\documentclass[fleqn,usenatbib]{mnras}

\usepackage{newtxtext,newtxmath}
\usepackage{multirow}
\usepackage[T1]{fontenc}
\DeclareRobustCommand{\VAN}[3]{#2}
\let\VANthebibliography\thebibliography
\def\thebibliography{\DeclareRobustCommand{\VAN}[3]{##3}\VANthebibliography}

\usepackage{float}

\usepackage{graphicx}  
\usepackage{amsmath}   
\usepackage[dvipsnames]{xcolor}
\title[Liller~1 according to APOGEE]{Is Liller~1 a building block of the Galactic bulge? - Evidence from APOGEE}

\newbox\grsign \setbox\grsign=\hbox{$>$} \newdimen\grdimen \grdimen=\ht\grsign
\newbox\simlessbox \newbox\simgreatbox
\setbox\simgreatbox=\hbox{\raise.5ex\hbox{$>$}\llap
     {\lower.5ex\hbox{$\sim$}}}\ht1=\grdimen\dp1=0pt
\setbox\simlessbox=\hbox{\raise.5ex\hbox{$<$}\llap
     {\lower.5ex\hbox{$\sim$}}}\ht2=\grdimen\dp2=0pt

\def\simless{\mathrel{\copy\simlessbox}}

\author[Liptrott et al.]{Anna Liptrott$^{1},$\thanks{E-mail: A.Liptrott@2020.ljmu.ac.uk} Ricardo P. Schiavon$^{1}$, Andrew C. Mason$^{1,8}$, Sebastian Kamann$^{1}$,
Borja Anguiano$^{6,7}$, \newauthor Roger E. Cohen$^{10}$, Jos\'e G. Fern\'andez-Trincado$^{2}$, Danny Horta$^{5}$, Steven R. Majewski$^{7}$, Dante Minniti$^{3,4}$, \newauthor David M. Nataf$^{9}$, Michael J. W. O'Connor$^{1}$, Dominic Wearne$^{1}$
\\
$^{1}$ Astrophysics Research Institute, Liverpool John Moores University, 146 Brownlow Hill, Liverpool L3 5RF, UK\\
$^{2}$ Instituto de Astronom\'ia, Universidad Cat\'olica del Norte, Av. Angamos 0610, Antofagasta, Chile \\
$^{3}$ Instituto de Astrof\'isica, Depto. de F\'isica y Astronom\'ia, Facultad de Ciencias Exactas, Universidad  Andres Bello, Av. Fern\'andez Concha 700, Santiago, Chile\\
$^{4}$ Vatican Observatory, V00120 Vatican City State, Italy \\
$^{5}$ Institute for Astronomy, University of Edinburgh, Royal Observatory,
Blackford Hill, Edinburgh EH9 3HJ, UK\\
$^{6}$ Centro de Estudios de F\'isica del Cosmos de Arag\'on (CEFCA), Plaza San Juan 1, 44001, Teruel, Spain\\
$^{7} $Department of Astronomy, University of Virginia, Charlottesville,
VA, 22904, USA\\
$^{8}$ Institute of Systems, Molecular, and Integrative Biology, University of Liverpool, Biosciences Building, Crown Street, Liverpool, Merseyside, L69 7BE\\
$^{9}$ Department of Physics \& Astronomy, University of Iowa, Iowa City, IA 52242, USA\\
$^{10}$ Department of Physics and Astronomy, Rutgers the State University of New Jersey, 136 Frelinghuysen Rd., Piscataway, NJ 08854, USA
}
\date{Accepted XXX. Received YYY; in original form ZZZ}
\pubyear{\the\year{}}

\begin{document}
\label{firstpage}
\pagerange{\pageref{firstpage}--\pageref{lastpage}}
\maketitle

\begin{abstract}
Liller~1 is a stellar system orbiting within the inner 0.8~kpc of the Galactic centre, characterised by a wide spread in age and metallicity, indicating a high mass. 
Liller~1 has been proposed to be a major contributor to the stellar mass of the Galactic bulge, yet its origin is subject to debate.
We employ Sloan Digital Sky Survey IV (SDSS-IV) data from the Apache Point Observatory Galactic Evolution Experiment (APOGEE) to test scenarios proposed to explain the nature of Liller~1. Using a random sampling technique, we contrast the chemical compositions of Liller~1 stellar members with those of the bulge, inner disc, outer disk and solar neighbourhood. The chemistry of Liller~1 deviates from that of the bulge population at the 2-3$\sigma$ level for $\alpha$-elements Mg, Si, and Ca. We conclude that the progenitor of Liller~1 was not a {\it major} contributor of stellar mass to the bulge. Furthermore, we find the abundance pattern of Liller~1 to deviate at the 2$\sigma$ level from that of inner disk stars, ruling out the cluster rejuvenation scenario. Finally, we find that Liller~1 is chemically distinct from solar and outer disc populations, suggesting that the progenitor of Liller~1 is unlikely to be an in-situ massive clump formed at high redshift, from disc gravitational instabilities, that migrated inwards and coalesced with others into the bulge. Finally, we suggest that Liller~1 is a {\it minor} contributor to the stellar mass of the inner Galaxy, possibly of extragalactic origin. 
\end{abstract}


\section{Introduction}
In the $\mathrm{\Lambda CDM}$ cosmological model, galaxy formation takes place through hierarchical merging, where low mass structures merge together to form the galaxies we observe today. 
The signatures of this process can be seen in the observed kinematics and chemical compositions of the stellar populations of the Milky Way (MW). 
Evidence of the accretion history of the MW has been accumulating over the years through the discovery of the Gaia-Enceladus-Sausage \citep[GES:][]{belokurov2018, Haywood2018, helmi2018, Mackereth2019}, the Helmi streams \citep[]{Helmi1999} and the Sagittarius dwarf spheroidal and streams \citep[]{ibata994, Majewski2013}. 
In particular, the inner few parsecs of the MW (a region referred to in the literature as the Galactic bulge) hosts complex stellar populations from the superposition of different Galactic components having distinct chemodynamical properties. Recently, \cite{Horta2021} provided evidence for the presence, within the inner 4~kpc of the MW halo, of the remnants of an early accretion of a massive satellite referred to as Heracles \citep[see also][]{Horta2024}.

The satellites involved in these mergers brought a plethora of globular clusters (GCs) to the MW \citep[e.g.,][]{massari2019, Myeong2019, kruijssen2019, 2023Montyb, Massari2023}. 
Consequently, as the oldest surviving stellar systems in the Galaxy, GCs are invaluable tracers of the early formation history of the MW. 
The bulge population of GCs span a wide range of properties \citep[e.g.,][]{barbuy1998, schiavon2017, geisler2021,geisler2023} with some of them hosting stellar populations with a wide range of ages and metallicities \citep[]{ferraro2021}. 
Two of these enigmatic systems, whose complex mix of stellar population still defy interpretation, are Terzan~5 \citep[]{ferraro2009} and Liller~1 \citep[]{ferraro2021}. The latter is one of the clusters located closest to the Galactic centre \citep[]{Baumgardt2019,Minniti2021} presenting a high central velocity dispersion \citep[]{Baumgardt2019}, indicative of a high mass and a complex
history. 

Photometric and spectroscopic studies of Terzan~5 \citep[]{ferraro2009,origlia2011, origlia_terzan_2013}, identified the presence of a multi-modal metallicity distribution with peaks at $\mathrm{[Fe/H] =-0.79, -0.25, and +0.3}$~dex. Analysis of Hubble Space Telescope-based proper motions showed the existence of two stellar populations widely separated in age \citep[$4.5$~Gyr and $12$~Gyr,][]{ferraro2016} with the younger population centrally segregated \citep[]{ferraro2009}. 
The distribution of Terzan~5 stars in the $\mathrm{[\alpha/Fe]}$ plane 
was revealed to resemble that of the Galactic bulge with \cite{ferraro2016} suggesting the progenitor mass of Terzan~5 to be as high as $\mathrm{10^{9-10} M_\odot}$. This body of evidence led to the hypothesis that Terzan~5 is the fossil remnant of a primordial building block of the MW bulge. 
These fossil fragments are hypothesised to be remnants of massive clumps formed at high redshift due to disc gravitational instabilities that migrated inwards \citep[]{immeli2004, Dekel2009} to coalesce to form the bulge. 

In addition to Terzan~5, Liller~1 was shown to be a complex bulge stellar system, hosting multi-age and multi-metallicity stellar populations with a present day mass exceeding $\mathrm{10^6~M_{\odot}}$ \citep[]{Lanzoni2010, ferraro2021}. 
Thus, the progenitor of Liller~1 was identified as another candidate building block of the Galactic bulge by \cite{ferraro2021}. Studies of the colour-magnitude diagrams of Liller~1 and Terzan~5 indicated the occurrence of prolonged star formation histories in both systems, with three star formation episodes each \citep[]{Dalessandro2022, crociati2023}. 
The best fit solution for the star formation history of Liller~1 is consistent with a main episode having started approximately 12-13~Gyrs ago, the second episode between 6 and 9~Gyrs ago, followed by an event starting 3~Gyrs ago \citep[]{Dalessandro2022}. 
Spectroscopic observations of Liller~1 revealed the existence of a bimodal metallicity distribution: a subsolar component at $\mathrm{[Fe/H] =-0.48}$~dex and a supersolar component at $\mathrm{[Fe/H] = +0.26}$~dex, with the latter found to be more centrally located \citep[]{crociati2023}, suggesting a self enrichment scenario. Spectroscopic follow ups identified a potential third stellar population within Liller~1 with an iron content and $\mathrm{[\alpha/Fe]}$ enhancement that are intermediate between those of the other two sub-populations \citep[]{Alvarez2024, Fanelli2024}. 

Further scenarios have been proposed to explain the nature of the two systems. \citet{Mckenzie2018, bastian2022} propose that these systems were globular clusters that underwent the accretion of gas from a giant molecular cloud (GMC) to form younger stellar generations. 
\citet{bastian2022} contend that dynamical friction makes the scenario that Liller~1 and Terzan~5 are bulge fossil fragments unlikely. Moreover, these authors discuss the complications regarding these systems losing mass as they contribute to the bulge stellar population, whilst remaining massive to self enrich to form new generations of stars. 
In addition, \citet{taylor2022} showed that the abundance pattern of Terzan~5 differs from that of bulge populations to a high statistical significance. The results of the analysis showed they differed at statistically significant levels leading to the conclusion that the progenitor of Terzan~5 was not an important contributor of stellar mass to the Galactic bulge.

In this paper, we test the hypothesis that the progenitor of Liller~1 was a major building block of the Galactic bulge through chemical tagging based on the precise elemental abundances for statistically significant samples from both systems. 
Furthermore, we compare the chemical composition of Liller~1 to inner disk, the Solar neighbourhood annulus and the outer disc population defined by Galactocentric distance 5-7kpc, 7-9kpc, and 9-11kpc, respectively, to test the origin of Liller~1 as a massive clump at high redshift that migrated inwards from the disk and to test the GC-GMC accretion scenario proposed to explain the nature of Liller~1.

The structure of this paper is as follows: in Section \ref{sec:datanmethod}, we describe the data used in this study and our procedure for selecting Liller~1 members and the field populations.
In Section \ref{sec:analysis}, we present the analysis of the chemical abundance patterns of Liller~1 and the bulge, Solar neighbourhood and inner disc fields. The results are presented in Section~\ref{sec:natureofliller1} and conclusions are summarised in Section \ref{sec:conclusions}.

\section{Data and Methods}
\label{sec:datanmethod}
\subsection{Data Source}

This paper utilises the seventeenth data release (DR17: \url{https://www.sdss4.org/dr17/}) of the SDSS-IV \citep[]{blanton2017} and APOGEE survey \citep[]{Majewski2017, abdurrouf2022}.
Proper motions are derived by cross-matching the catalogue with the third early data release of the Gaia survey \citep[eDR3;][]{prusti2016, brown2021}. 
The APOGEE DR17 catalogue employed in this work (suffix \textsc{synspec-rev1}) comprises high precision chemical abundances and line of sight velocities for $\sim$ 700,000 stars within the Milky Way, a number of its satellites and globular clusters.
The elemental abundances and radial velocities are obtained from the analysis of high resolution near infrared spectra observed with the twin multi-fibre spectrographs \citep[]{wilson2019} attached to the 2.5~m Sloan Foundation Telescope at the Apache Point Observatory \citep[APO;][]{gunn2006} in New Mexico, USA, and the 2.5~m du Pont Telescope \citep[]{bowen1973} at the Las Campanas Observatory (LCO) in Chile, enabling coverage of both hemispheres.
The description of the target selection can be found in \cite{zasowski2017}, \cite{beaton2021}, and \cite{santana2021}.
From the NIR spectra and stellar parameters and the detailed chemical compositions are determined by the APOGEE Stellar Parameter and Chemical Abundances Pipeline \citep[ASPCAP;][]{holtzman2015, perez_aspcap_2016}.
Further information regarding the APOGEE survey, data reduction, and abundance pipelines are found in \cite{Majewski2017}, \cite{holtzman2018}, and \cite{jonsson2020} respectively.
The stellar distances for the APOGEE DR17 catalogue are obtained using a retrained AstroNN neural network software \citep[for a full description see][]{leung2019b, leung2019a}, which predicts stellar luminosity from spectra using a training set comprised of stars with both Gaia eDR3 parallax measurements and APOGEE DR17 spectra.

\subsection{Parent sample}
To only consider stars with spectral parameters determined with confidence and reliable elemental abundances we limit the sample to those with STARFLAG $=$ 0 (indicating no issues in the spectra impacting the fit) and whose combined spectra have S/N $>$ 70. The sample is limited to stars whose combined spectra have S/N $>$ 70 and STARFLAG $=$ 0. ASPCAP flags are considered on an individual basis through investigating the spectra and lines of interest. 

Stars in the parent sample are further selected based on the following criteria: those with reliable APOGEE stellar surface parameters: 3000 $\mathrm{< T_{eff} <}$ 6000~K and $\log g < $ 3.6~\rm{dex}. At high $\mathrm{T_{eff}}$ abundances become uncertain due to the weakness of absorption lines. The low $\mathrm{T_{eff}}$ limit is imposed as the model atmospheres become uncertain, so that abundances are affected by important systematic effects. We further remove all likely binary/pulsating stars from the sample (i.e. $\mathrm{VSCATTER < 2~km s^{-1}}$).

\subsection{Galactic bulge sample} 
The Galactic bulge sample is defined on the basis of a simple spherical spatial cut: Galactocentric distance $\mathrm{R_{GC} < 4~kpc}$.
The Galactocentric distance, $\mathrm{R_{GC}}$ for each star was determined using Galactic longitude $l$, latitude $b$, the AstroNN distance $d$, distance error ${d_{\rm err}}$, and assuming a distance of 8~kpc between the Sun and the Galactic Centre \citep[]{utkin2020}. Further cuts are applied to remove stars with large distance errors - those with fractional distance uncertainty $\mathrm{\sigma_d > 0.2}$. Contamination of the bulge sample by stars with apocentres beyond 4~kpc is limited to seven stars. Stars for which the abundances in Mg, C, N, O and Si could not be determined by ASPCAP were removed. In addition, GC stars belonging to the APOGEE GC Value Added Catalogue \citep[]{Schiavon2024} are removed. Application of these selection criteria resulted in a bulge sample of 20,046 stars with reliable elemental abundances and distances.

\subsection{Galactic disc sample}
\label{sec:disksample}
Following the procedure defining the Galactic bulge sample, spacial cuts are applied to the parent sample to derive three subsets of Galactic disc stars. The three subsets of disc field stars are defined as follows:

\begin{itemize}
    \item Inner Disk: $\mathrm{5 < R_{Gc} < 7~kpc}$ and $\mathrm{|z| < 2.0~kpc}$
    \item Solar Neighbourhood: $\mathrm{7 < R_{Gc} < 9~kpc}$ and $\mathrm{|z| < 2.0~kpc}$
    \item Outer Disk: $\mathrm{9 < R_{Gc} < 11~kpc}$ and $\mathrm{|z| < 2.0~kpc}$
\end{itemize}

\subsection{Liller~1 candidates}
\label{sec:candidates}

Liller~1 candidate members were determined on the basis of angular distance from the cluster centre, radial velocity and proper motion. The central coordinates for Liller~1, $\mathrm{\alpha_{Lil1} = 17^h33^m24^s.50}$ and $\mathrm{\delta_{Lil1} = -33^{\circ}46'45''}$ (J2000) are adopted from \cite{harris_new_2010}. 
The relevant values used to describe Liller~1 are summarised in Table \ref{tab:Liller1_selection_criteria}. Stars are considered candidate members of Liller~1 if they lie within the cluster's Jacoby radius and have radial velocities differing from the mean radial velocity \citep[$\mathrm{<v_r> = 60.35 kms^{-1}}$;][]{Baumgardt2019} by less than two times the cluster's velocity dispersion.

Additional criteria are based on the Gaia eDR3 \citep[]{brown2021,lindegren2021} proper motions. Liller~1 candidate members are determined as those whose proper motions do not deviate from the mean value of the cluster \citep[]{baumgardt2021} by more than five times the cluster mean proper motion dispersion, $\mathrm{\sigma_{PM}}$ \citep[]{Schiavon2024}.

To further refine our selection of candidate Liller~1 members, we implement the Bayesian maximum likelihood approach that takes into account radial velocities and projected distances of the stars to assess the probability of membership, presented in \cite{walker2009} and \cite{kamann2018}: The line of sight velocities distribution is modelled as the superposition of two Gaussians, where the cluster component is described through a Plummer density and velocity dispersion profile \citep[]{plummer1911}. The field component is modelled assuming a constant spatial density and velocity dispersion profile. The model parameters are optimised using \textsc{emcee} \citep[]{Foreman2013}, a Python implementation of the affine-invariant Markov Chain Monte Carlo (MCMC) ensemble sampler, and membership probabilities assigned to each star. 

\begin{figure}
    \includegraphics[width=1\columnwidth]{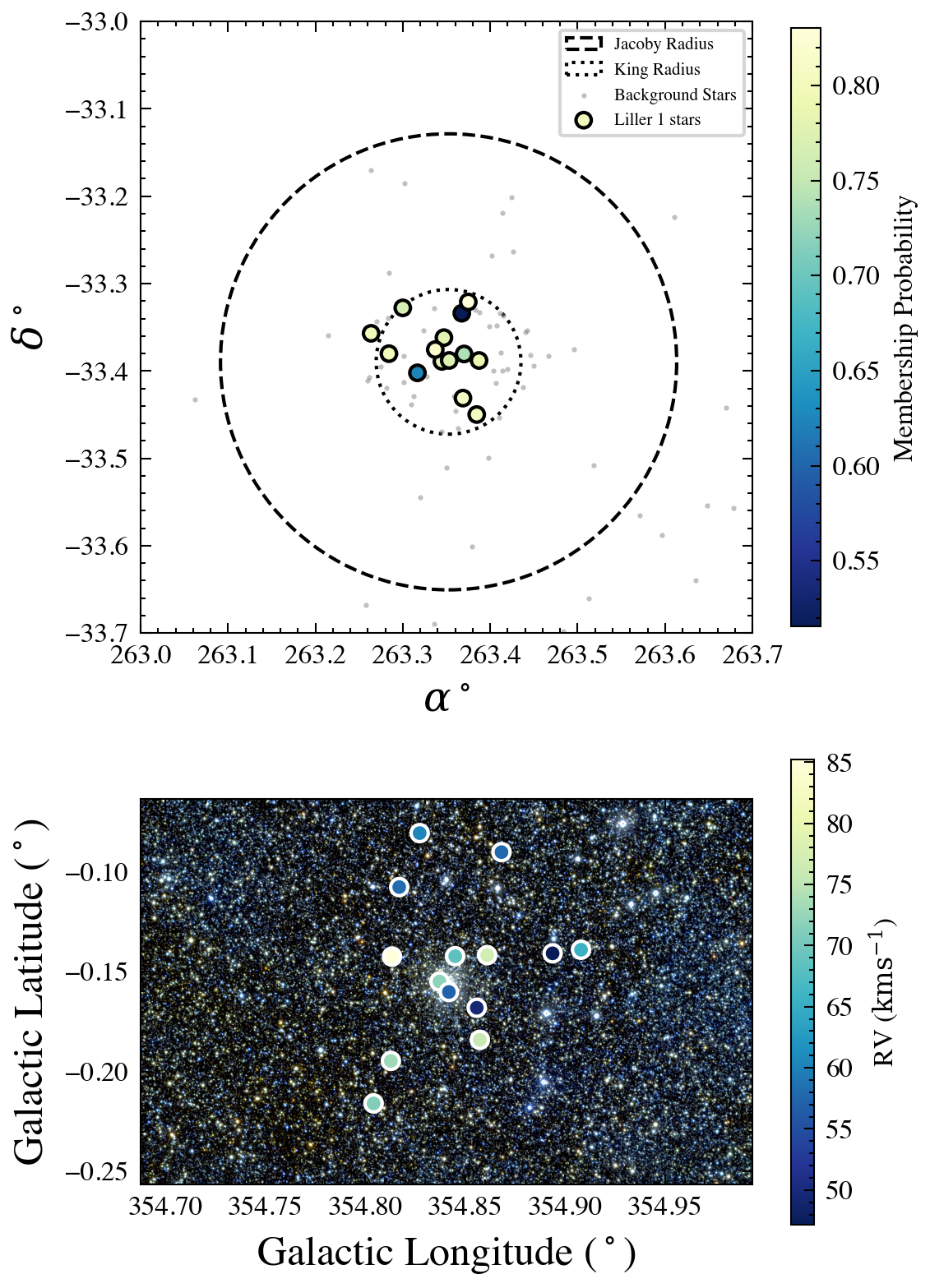}
    \caption{Top panel: Right ascension $\alpha$ and declination $\delta$ (in degrees) of the Liller~1 stars colour coded by probability of cluster membership
    The cluster Jacoby radius $r_j$ \citep[][black dashed line]{baumgardt2021} displayed as a reference. The smaller tidal radius from \citep[]{saracino2015} is shown with the dotted line. All but one of the stars in our sample of Liller~1 candidates are contained within the tidal radius. Bottom panel. Near-IR colour image from the VVV Survey made using the JHKs filters \citep[]{minniti_vista_2010} covering approximately $18^{\prime} \times 12^{\prime}$, oriented along Galactic coordinates, with longitude increasing to the left and latitude increasing upwards. Overlaid are the Liller~1 stars colour-coded by radial velocity.}
    \label{fig:ra_dec}
\end{figure}

The candidate stars, colour-coded by membership probability, are plotted in Figure \ref{fig:ra_dec} with APOGEE DR17 background stars shown. Liller~1 stars are considered certified cluster members above the threshold of a probability of membership, $\mathrm{p > 0.5}$, resulting in fourteen Liller~1 stars. These stars are listed in Table~\ref{tab:Liller1Candidates}, together with APOGEE {\it calibrated} atmospheric parameters, S/N ratio of the combined spectrum, radial velocity, projected distance from the centre of Liller~1, and the abundances of key elements. 

{  The uncertainties on stellar parameters in DR17, see Table \ref{tab:Liller1Candidates}, are estimated from the difference in repeat observations of stars, as such these only represent the random uncertainties and do not reflect the systematic uncertainties which are expected to be larger. We refer the reader to Tables 12 and 13 in \citet{jonsson2020} for approximate systematic uncertainties on abundances and stellar parameters.}

\begin{table*}
    \centering
    \caption{Summary of the parameters used to select candidate Liller~1 stars, including cluster mean RA $\mathrm{\alpha_{Lil 1}}$, and Dec $\mathrm{\delta_{Lil 1}}$, Jacoby radius $r_j$ in arcminutes, mean heliocentric cluster radial velocity $\mathrm{RV_{Lil 1}}$ and central velocity dispersion $\sigma_{RV}$ from \citet{Baumgardt2018} in $\mathrm{kms^{-1}}$. RA proper motion $\mu_\alpha \cos(\delta)$ and Dec proper motion $\mu_\delta$ in $\mathrm{mas yr^{-1}}$, mean proper motion dispersion in $\mathrm{mas yr^{-1}}$, and galactocentric cluster distance $\mathrm{R_{GC}}$ in kpc.} 
    \begin{tabular}{clllllllr}
        \hline
         $\alpha^\circ_{Lil 1}$ & $\delta^\circ_{Lil 1}$ & $r'_J$ & $RV $ & $\sigma_{RV}$ & $\mu_\alpha \cos(\delta)$ & $\mu_\delta$ & $\sigma_{PM}$ & $R_{GC}$ \\
         \hline
         263.35233 & -33.38956 & 15.66 & 60.36 $\mathrm{kms^{-1}}$ & 23.2 $\mathrm{kms^{-1}}$& -5.183 $\mathrm{mas yr^{-1}}$,& -6.979 $\mathrm{mas yr^{-1}}$,& 0.67 $\mathrm{mas yr^{-1}}$ & 0.74 kpc\\
         \hline
    \end{tabular}    
\label{tab:Liller1_selection_criteria}
\end{table*}

\begin{table*}
\caption{Summary of the sample of Liller~1 stars including the calibrated stellar parameters and abundances from APOGEE. 2M17331198-3319400 does not have an associated [Ca/Fe] abundance from APOGEE. The nominal DR17 $T_{\rm eff}$, $\log g$, and abundance errors are smaller than 6~K, 0.03, and 0.02~dex. They are based on measurements on repeat observations of a set of stars.  Being immune to systematic effects, reflect precision, rather than real uncertainties in the measurements.}
\resizebox{\textwidth}{!}{%
\begin{tabular}{@{}lllllllllllll@{}}
\hline
APOGEE ID & $T_\mathrm{eff} (K)$ & $\log g$ & SNR & RV ($\mathrm{kms^{-1}}$)& $r_j'$ & {[}Fe/H{]} & {[}C/Fe{]} & {[}N/Fe{]} & {[}O/Fe{]} & {[}Mg/Fe{]} & {[}Si/Fe{]} & {[}Ca/Fe{]}\\
\hline
2M17330328-3321249 & 3428 & 0.48 & 182 & +60.5 & 4.85 & --0.08 & +0.01 & +0.07 & +0.08 & +0.04 & --0.04 & --0.07 \\
2M17330818-3322487 & 3469 & 0.79 & 181 & +58.2 & 3.46 & +0.28 & +0.04 & +0.14 & +0.06 & --0.00 & --0.06 & --0.14 \\
2M17331198-3319400 & 3268 & 0.43 & 225 & +57.6 & 4.54 & +0.21 & +0.04 & +0.15 & +0.06 & --0.01 & --0.05 & \\
2M17331600-3324077 & 3565 & 1.19 & 128 & +85.2 & 1.94 & +0.27 & +0.02 & +0.28 & +0.07 & --0.05 & --0.04 & --0.13 \\
2M17332272-3323206 & 3399 & 0.72 & 198 & +72.0 & 0.38 & +0.21 & +0.02 & +0.18 & +0.06 & --0.04 & --0.02 & --0.10\\
2M17332328-3321427 & 3444 & 0.88 & 189 & +77.0 & 1.68 & +0.38 & +0.06 & +0.18 & +0.05 & +0.00 & --0.06 & --0.09 \\
2M17332472-3323166 & 3516 & 0.84 & 214 & +57.2 & 0.10 & --0.08 & +0.02 & +0.07 & +0.05 & --0.04 & --0.04 & --0.14 \\
2M17332817-3320020 & 3404 & 0.60 & 142 & +47.2 & 3.42 & +0.17 & +0.03 & +0.12 & +0.06 & +0.02 & --0.07 & --0.17 \\
2M17332850-3325518 & 3739 & 0.82 & 139 & +73.0 & 2.62 & --0.58 & +0.18 & +0.21 & +0.33 & +0.31 & +0.18 & +0.04 \\
2M17332881-3322499 & 3703 & 0.74 & 162 & +49.4 & 1.04 & --0.62 & +0.15 & +0.27 & +0.32 & +0.30 & +0.18 & +0.05 \\
2M17332996-3319156 & 3667 & 1.11 & 131 & +65.6 & 4.26 & +0.23 & +0.04 & +0.28 & +0.05 & --0.00 & --0.04 & --0.14 \\
2M17333228-3326596 & 3504 & 0.57 & 143 & +71.2 & 3.96 & --0.18 & +0.01 & +0.02 & +0.06 & +0.01 & --0.05 & --0.13 \\
2M17333292-3323168 & 3618 & 0.71 & 169 & +75.9 & 1.75 & --0.59 & +0.22 & +0.27 & +0.33 & +0.30 & +0.19 & +0.02 \\
2M17332090-3322320 & 3732 & 1.50 & 218 & +68.5 & 1.13 & --0.59 & --0.16 & +0.82 & +0.11 & +0.12 & +0.05 & +0.12 \\
\hline
\end{tabular}%
}
\label{tab:Liller1Candidates}
\end{table*}

Measurements of the tidal radii of GCs are uncertain---a problem furthered in the crowded Galactic bulge. \cite{saracino2015} obtains a value of $4.97$' for the tidal radius of Liller~1, by fitting a King model to GEMINI/GeMS NIR observations. The much larger Jacoby radius from \cite{baumgardt2021} results from a dynamical calculation matching the cluster’s stellar density and velocity dispersion profiles (derived from a combination of literature sources and Gaia eDR3). The Jacoby radius depends on the cluster’s mass and orbit.
By definition, the Jacoby radius and King tidal radius do not necessarily agree, as clusters do not follow a King profile at distances of the order of the Jacoby radius. However, thirteen of the candidate members of Liller~1 in this study are within the \cite{saracino2015} tidal radius so that the adoption of the larger Jacoby radius has negligible impact on our results.

\subsection{Stars in common with the \texorpdfstring{\citet{Fanelli2024}}{Fanelli et al. (2024)} analysis}

\begin{figure}
    \includegraphics[width=0.8\columnwidth]{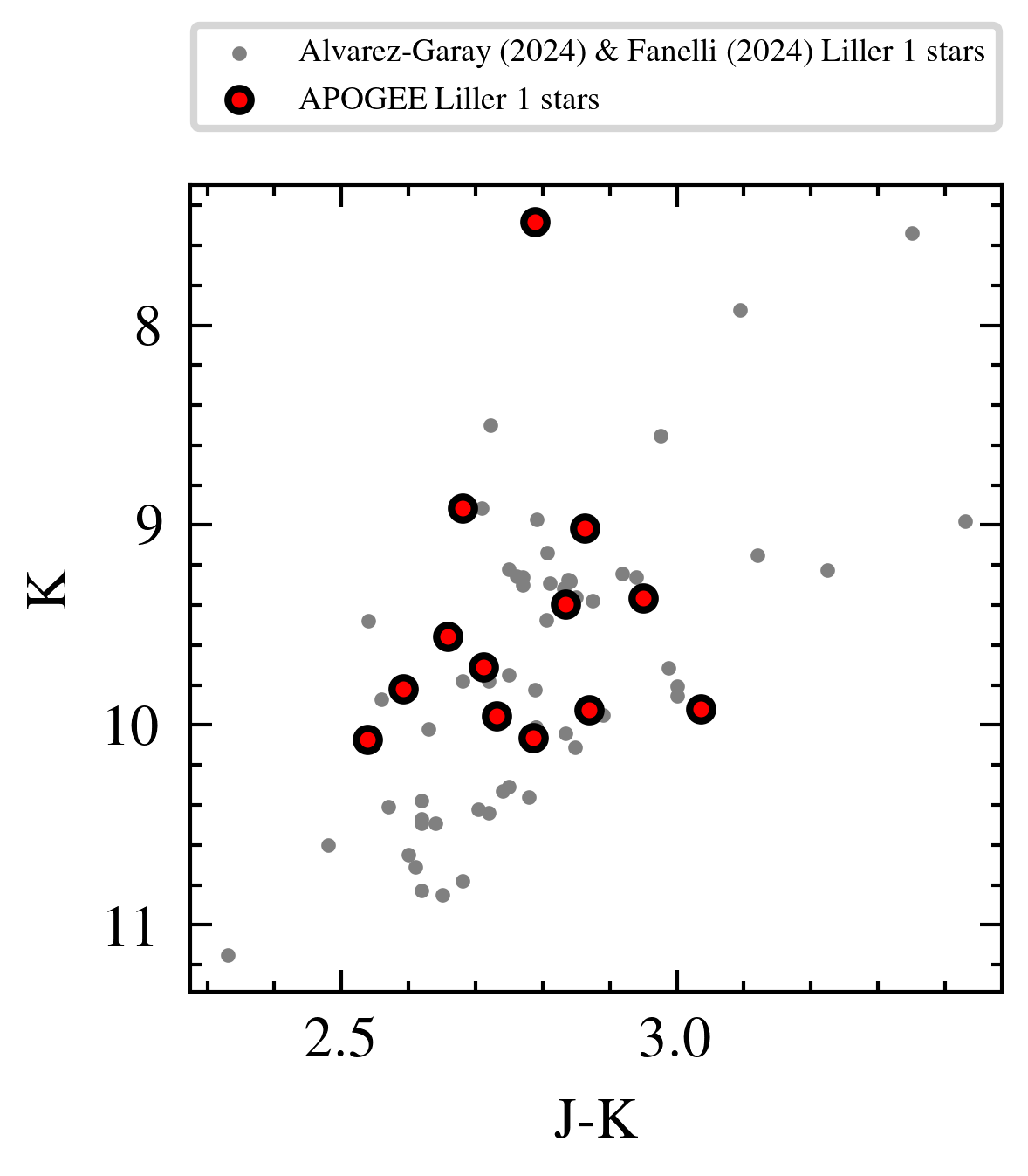}
    \caption{(J)-(J-K) colour magnitude diagram of Liller~1 stars in this paper in red with those from \citet{Alvarez2024,Fanelli2024} plotted in grey. {  For a Liller~1 sample that extends further down the red giant branch we refer the reader to \citet{Valenti2010} and \citet{ferraro2021}}.}
    \label{fig:cmd}
\end{figure}

In a recent paper, \citet{Fanelli2024} reported the analysis of Keck II/NIRSPEC spectra for a sample of Liller~1 stars. Figure~\ref{fig:cmd} shows that their sample and ours occupy the same locus in the NIR colour-magnitude diagram. 
Three stars in the final APOGEE Liller~1 sample are in common with those from the \citet{Fanelli2024}. Two of these are reported by those authors to have nitrogen abundances exceeding the threshold $\mathrm{[N/Fe] = +0.5}$, in disagreement with the APOGEE abundances by 0.5-0.6 dex. 
The IDs of these three stars, alongside their respective parameters are summarised in Table~\ref{tab:comp}. Also included are the {\it spectroscopic} APOGEE stellar parameters, i.e., those corresponding to the uncalibrated ASPCAP output. 
They are the parameters that lead to the best fit to the observed APOGEE spectrum. There are important discrepancies in $T_{\rm eff}$ (between 76 and 167~K) and $\log g$ (0.25 to 0.42~dex), which are just about within the nominal errors of the two sets of measurements. Among the elemental abundances, the greatest differences are seen in [N/H]. For the two stars referred above, \cite{Fanelli2024} find values higher than those by APOGEE by $\approx +0.6$~dex.
Since that would place both stars in the 2G locus of the abundance space, it is important that we investigate the root cause of these discrepancies, which we do in Section~\ref{sec:synthesis}.  However, in the next sub-section we first briefly describe the method employed in the derivation of APOGEE abundances, for completeness.

\subsubsection{A note on the ASPCAP analysis of cool stars}

The spectroscopic analysis of M giants is often viewed with  apprehension on account of various well known difficulties.  On the theory side, model atmospheres are affected by incomplete and/or inaccurate molecular line lists, as well as uncertainties in the treatment of convection, mixing, pulsations, variability, mass loss, and dust formation, non-LTE effects, and the treatment of geometry (i.e., 1D vs 3D, plane-parallel vs spherically symmetrical models).  
Some of these problems, such as line list limitations, treatment of geometry, and non-LTE effects also impact the quality of the spectrum synthesis \citep[e.g.,][]{plez2008}. On the observational side, the main difficulties are associated with the treatment of the observed spectrum, in view of the virtual absence of clear regions of continuum, particularly at moderate resolution. 
In view of this, it is fitting that we provide a brief description of the methods adopted in the spectral analysis by APOGEE, while referring the reader to the relevant literature for a more detailed account.

A cornerstone of the procedure is the generation of a huge library of synthetic spectra based on state-of-the-art input physics and radiative transfer modelling.  APOGEE produced abundance catalogues based on libraries of several flavours.  
This paper analyses abundances generated by optimisation of observed spectra against a synthetic library calculated with the \textsc{synspec} radiative transfer code (suffix \textsc{synspec-rev1}), so we focus on that specific library.
The model atmospheres adopted are those from the MARCS family, calculated under the plane-parallel approximation \citep{Gustafsson2008}. The library includes several subgrids. The one that is relevant to this work spans:
\begin{itemize}
    \item $\rm{3000 \leq T_{eff} \leq 8000K}$ 
    \item ${\rm-0.5 \leq log g \leq 5}$
    \item ${\rm -2.5 \leq [M/H] \leq 1.0}$
    \item ${\rm -0.5 \leq [\upalpha/Fe] \leq 1.0}$
    \item ${\rm -0.5 \leq [C/M] \leq 1.0}$
    \item Five values of $\nu_{\rm mic}$ (0.3, 0.6, 1.2, 2.4, 4.8~km~${\rm s^{-1}}$)
    \item Seven steps of ${\rm v \sin i}$ (1.5, 3, 6, 12, 24, 48, 96~km~${\rm s^{-1}}$)
\end{itemize}

The atomic and atomic and molecular line lists adopted are described by \cite{shetrone2015} and updated by \cite{Smith2021}. ASPCAP employs a two-step process to extract abundances. First, the code \textsc{FERRE} \citep{allendeprieto2006} performs interpolation within the above library, to determine stellar atmospheric parameters by fitting the entire APOGEE spectrum. 
The interpolation is done in eight dimensions: effective temperature, surface gravity, microturbulence, macroturbulence/rotation, overall metallicity, and the abundances of $\rm{\alpha}$-elements, C, and N. 
To speed up the calculation, \textsc{FERRE} resorts to principal component analysis, so that the interpolation is not done in spectral, but rather in principal component coefficient space.

Including the C, N and $\alpha$ abundances in the initial fitting process is important for cool, metal-rich stars, as their H-band spectra contain prominent molecular features due to CN, CO, and OH, which must be accurately modelled for the determination of reliable stellar parameters. 
This consideration also extends to continuum normalisation, where the smoothed residual between the observed and synthetic spectra is used to apply a small adjustment, accounting for the pseudo-continuum features introduced by these molecular bands.

Following the above coarse characterisation, the spectra are continuum normalised; this is a necessary step preceding the determination of detailed elemental abundances. Each observed spectrum is first divided by a heavily smoothed ratio between itself and the best match resulting from the coarse characterization. 

Starting from the above parameters, the next step is to determine individual elemental abundances. \textsc{FERRE} is used to fit narrow spectral windows chosen for their sensitivity to the abundances of each element. Each spectral window is weighted according to factors such as line strength and blending, with multiple windows contributing to the abundance determination of any given element. 

The above summary follows broadly the much more detailed accounts by \cite{holtzman2018} and \cite{jonsson2020}, which refer to the 14$^{\rm th}$ and 16$^{\rm th}$ data releases, respectively. There are relatively few differences in the data reduction and analysis between the latter and DR17, upon which this work is based.
The most important novelty in DR17 is the adoption of the \textsc{synspec} spectrum synthesis code, which implements NLTE calculations for Na, Mg, K, and Ca, at the expense of being limited to plane-parallel models. The differences between that and results obtained using \textsc{turbospectrum} \citep{plez2012} in DR16 are subtle and are mentioned briefly in Section~\ref{sec:quantify}. For more details on the differences between DR17 and DR16, see \cite{abdurrouf2022}.

All in all, there is no question that the complexities involved in the spectral analysis of cool stars inspire caution. However, we stress that the methods adopted in the analysis of APOGEE spectra are designed with an eye towards guaranteeing great consistency and reproducibility of the results. 
It is very difficult to rid the final numbers from systematic effects stemming from limitations in the spectrum synthesis code, the line lists, and the input physics of the models employed in the analysis. Nevertheless, the APOGEE pipeline relies on the consistent application of automatic methods to high quality spectra of similar stars, which results in an extremely high degree of precision in the resulting stellar parameters and elemental abundances. Therein lies the strength of using highly homogeneous samples for statistically robust comparisons, such as those of \cite{taylor2022} and this work.

\subsubsection{Spectrum Synthesis}
\label{sec:synthesis}

{In order to investigate the origin of
the abundance discrepancies with the work of \cite{Fanelli2024}, we calculate synthetic spectra based on their atmospheric parameters and chemical compositions to evaluate how well they reproduce the very high S/N, well calibrated, APOGEE spectra. To verify the quality of our spectrum synthesis, we also compare APOGEE spectra with calculations generated adopting the APOGEE stellar parameters and chemical compositions, finding a very good match, as expected.

For context, \cite{Fanelli2024} employ the \textsc{turbospectrum} \citep[]{Alvarez1998, Gerber2023} spectrum synthesis code, combined with the MARCS model atmospheres \citep[]{Gustafsson2008}. In addition, their synthesis is based on the VALD3 compilation of atomic lines \cite{Ryabchikova2015}, and a list of molecular lines made available publicly by B.~Plez.  Their abundances are anchored on the \cite{Magg2022} values for the Sun.

Our discussion is also based on calculations performed with \textsc{turbospectrum} and the MARCS model atmospheres. 
In contrast, we adopt the APOGEE DR17 line list \citep[]{Smith2021} and solar reference abundances from \citet{Grevesse2007}. We explored the impact of variations in these inputs, as reported towards the end of this sub-section.
To facilitate the comparison with the combined APOGEE stellar spectrum, the synthesised spectra are interpolated onto the \textsc{apStar} wavelength grid and convolved with the APOGEE DR12 line spread function (LSF), including macroturbulent broadening. Each spectrum is normalised by fitting a low order polynomial to the top 10\% of flux values for the separate regions of interest.

For each star, a number of MARCS model atmospheres are selected bracketing its values of effective temperature ($T_{\rm eff}$), surface gravity ($\log g$), and metallicity ([Fe/H]), for both the APOGEE and \cite{Fanelli2024} sets of stellar parameters. These model atmospheres are used to generate a grid of synthetic spectra covering the full range of stellar parameters and chemical compositions within the bracketing values. 
For each set of stellar parameters, the comparison spectra are generated via interpolation within the grid of synthetic spectra through a radial basis function interpolation method. 
The APOGEE and \cite{Fanelli2024} sets of stellar parameters are listed in Table \ref{tab:comp}. As the uncertainties in DR17 are small, for the reason discussed in Section~\ref{sec:candidates}, we adopt the DR16 uncertainties for the comparisons and for defining the shaded regions around the synthetic spectra, which represent the parameter uncertainty space. 
The DR16 stellar parameter uncertainties for $\rm T_{eff}$ and $\rm \log g,$ are derived from the scatter about the calibration relations; however, these values still likely underestimate the true systematic uncertainties.}

\begin{table*}
\caption{Liller~1 stars in common and the derived parameters for APOGEE (top row for each star) and those from \citet{Fanelli2024}. The errors adopted for the APOGEE parameters are from DR16 (see text).}
    \centering
    \begin{tabular}{clllllllr}
        \hline
APOGEE/F24 ID & Parameters & {$\rm{T_{eff}}$} (K) & {\rm{$\log{g}$}} & [Fe/H] & [C/H] & [N/H] & [O/H] &  ${\rm \xi_t (kms^{-1})}$ \\
\hline
\multirow{3}{*}{2M17332272-3323206/14} & APOGEE & 3324$\pm$54 & 0.92$\pm0.04 $& +0.21$\pm$0.01 & +0.22$\pm$0.01 & +0.39$\pm$0.01 & +0.26$\pm$0.01 & 2.26\\
& APOGEE (calibrated) & 3399 & 0.72 & \multicolumn{5}{c}{}\\
                 & F24 & 3400$\pm$100 & 0.5$\pm$0.3 & +0.24$\pm$0.03 & -0.15$\pm$0.07 & +0.64$\pm$0.12 & +0.26$\pm$0.04 & 2.0\\
                 & $\Delta_{\rm{APOGEE - F24}}$ & -76 & +0.42 & -0.03 & +0.37 & -0.25 & 0.00 & +0.26\\
\hline
\multirow{3}{*}{2M17332881-3322499/70} & APOGEE & 3567$\pm$69 & 0.7 $\pm$0.05& -0.62$\pm$0.01 & -0.47$\pm$0.01 & -0.34$\pm$0.01 & -0.30$\pm$0.01 & 2.36 \\
& APOGEE (calibrated) & 3703 & 0.74 & \multicolumn{5}{c}{}\\
                 & F24 & 3400$\pm$100 & 0.5$\pm$0.3 & -0.48$\pm$0.03 & -0.62$\pm$0.05 & +0.21$\pm$0.10 & -0.11$\pm$0.10 & 2.0\\
                 & $\Delta_{\rm{APOGEE - F24}}$ & +167 & +0.26 & -0.14 & +0.15 & -0.55 & -0.19 & +0.36\\
\hline
\multirow{3}{*}{2M17333292-3323168/67} & APOGEE & 3484$\pm$66 & 0.75$\pm$0.05 & -0.59$\pm$0.01 & -0.37$\pm$0.01 & -0.32$\pm$0.01 & -0.26$\pm$0.01 & 1.82 \\
& APOGEE (calibrated) & 3618 & 0.71 & \multicolumn{5}{c}{}\\
                 & F24 & 3400$\pm$100 & 0.5$\pm$0.3 & -0.34$\pm$0.04 & -0.48$\pm$0.05 & +0.29$\pm$0.10 & +0.13$\pm$0.10 & 2.0\\
                 & $\Delta_{\rm{APOGEE - F24}}$ & +84 & +0.25 & -0.25 & +0.11 & -0.61 & -0.39 & -0.18\\
                                
\hline
\end{tabular}%
\label{tab:comp}
\end{table*}

\begin{figure*}
    \includegraphics[width=1\textwidth]{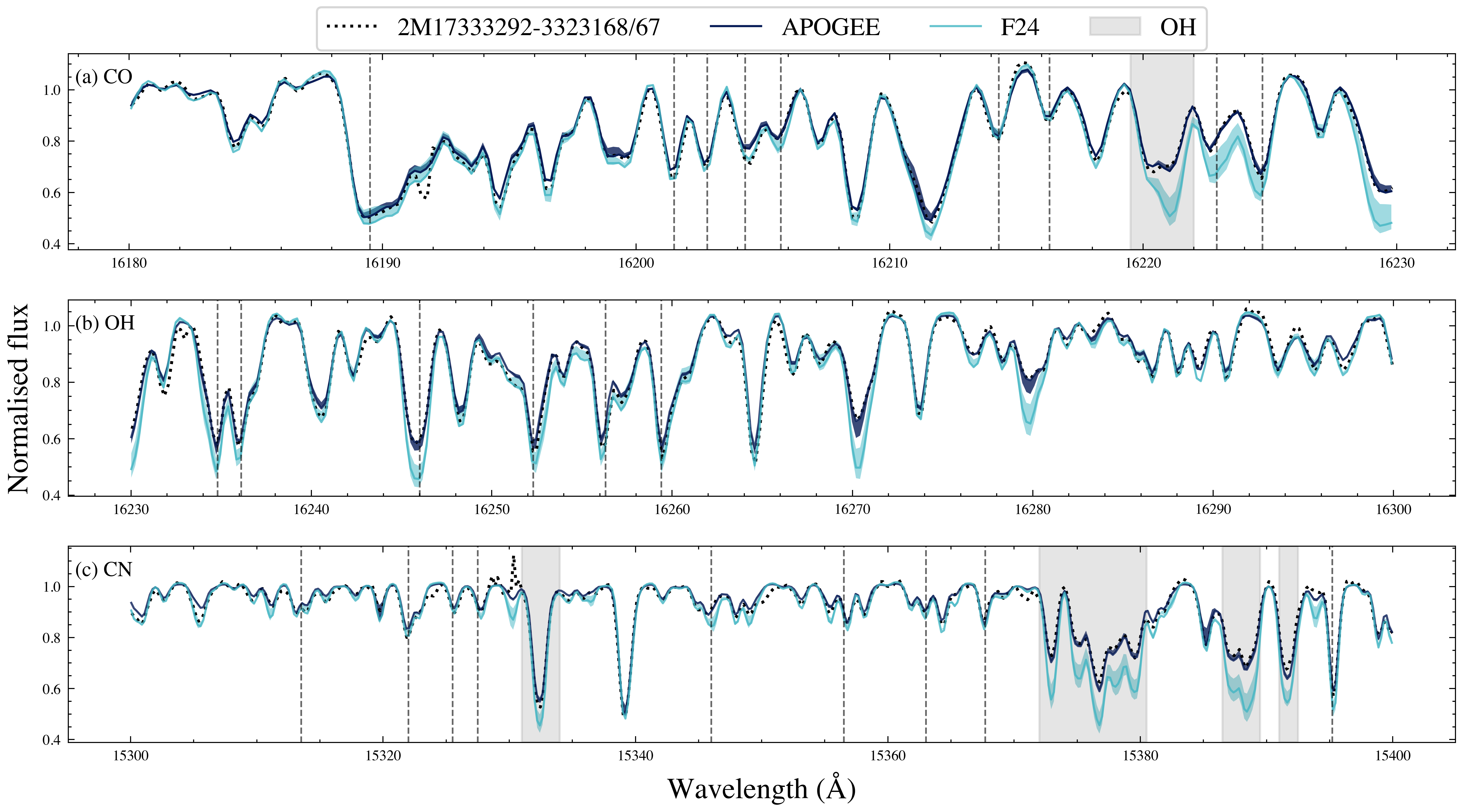}
    \caption{Combined APOGEE spectra for star 2M17333292-3323168/67 in black (dotted line). The synthetic spectrum for the F24 parameters is shown in light blue, and the APOGEE parameters spectrum is shown in navy. The shaded regions around each spectrum indicates the error. (a), (b), (c): comparison between both spectra and the observed spectrum for the CO, OH, CN molecular bands respectively. The lines are indicated with the grey dashed lines. The grey shaded regions indicate further OH bands.}
    \label{fig:spec}
\end{figure*}

The resulting normalised spectra are shown in three panels in Figure \ref{fig:spec} for one of the stars for which the \citet{Fanelli2024} and APOGEE N abundances differ substantially. The panels display both synthetic spectra and the combined spectrum for star 2M17333292-3323168 within spectral regions covering predominantly lines due to CN, OH, and CO, whose positions are indicated.

Examination of Figure~\ref{fig:spec} elucidates the differences in stellar parameters and chemical compositions between APOGEE and \citet{Fanelli2024}. Overall the two synthetic spectra are a good match to the observed APOGEE spectrum, except that the strengths of OH lines are systematically overestimated when the \cite{Fanelli2024} parameters are adopted. This mismatch results from the fact that [O/H] is higher in the latter set of parameters by 0.39~dex. Interestingly, the synthetic spectrum calculated with the \cite{Fanelli2024} parameters is a relatively good match to CO and CN lines. That is not surprising. It is a consequence of the fact that [C/H] according to \cite{Fanelli2024} is lower than the APOGEE value by 0.11~dex, so that CO lines are well matched. The latter implies then that [N/H] has to be higher by 0.61~dex so that CN lines are not under predicted. 

The differences in $T_{\rm eff}$, $\log g$, and [Fe/H] also affect both  molecular dissociation equilibrium and opacities. The $T_{\rm eff}$ according to \citet{Fanelli2024} is slightly cooler, and [Fe/H] slightly higher, which leads to stronger molecular lines. Conversely, their $\log g$ is lower, which impacts positively the strengths of CN and CO lines, and negatively those of OH.
All in all, there most likely is a trade off between the effects of all these parameters on molecular line strengths. Nevertheless the inference that the O abundances by \citet{Fanelli2024} are overestimated by up to $\approx+0.4$~dex, leading up to an underestimate in C and an overestimate in N seems solid.

{  It is important that we check that our results are immune to systematic effects arising from the adopted line list and solar reference abundances. 
With that in mind, the synthesis is repeated by adopting the \citet{Fanelli2024} parameters using instead the list of atomic lines from the VALD3 compilation, and the variations of molecular line lists summarised in Table \ref{tab:molecularlinelists}.  
In addition, we also adopted the reference solar abundances from \citet[]{Magg2022}. 
The sources of molecular lines in Table \ref{tab:molecularlinelists} includes those from APOGEE, VALD3 and those from the Plez molecular line list for CO, CN, OH. 
The other molecular lines are from VALD3, the APOGEE DR14 line list or the APOGEE DR17 line list in order to ensure immunity from systematics arising from the choice of line lists.}

\begin{table}
\centering
\caption{The sources of molecular line lists for CO, OH, and CN used for the synthetic spectrum in Figure \ref{fig:valdmaggf24}.}
\begin{tabular}{|p{2cm}|p{4cm}|}
\hline
\textbf{Molecule} & \textbf{Line List Source(s)} \\
\hline
CO & 
    \begin{tabular}[t]{@{}l@{}}
        \citet{Goorvitch1994} \\
        \citet{li2015}
    \end{tabular} \\
\hline
OH & 
    \begin{tabular}[t]{@{}l@{}}
        \citet{Goldman1998} \\
        \citet{Kurucz2007} \\
        \citet{brooke2016}
    \end{tabular} \\
\hline
CN & 
    \begin{tabular}[t]{@{}l@{}}
        \citet{Sneden2014} \\
        \citet{brooke2016}
    \end{tabular} \\
\hline
\end{tabular}
\label{tab:molecularlinelists}
\end{table}



\begin{figure*}
    \includegraphics[width=1\textwidth]{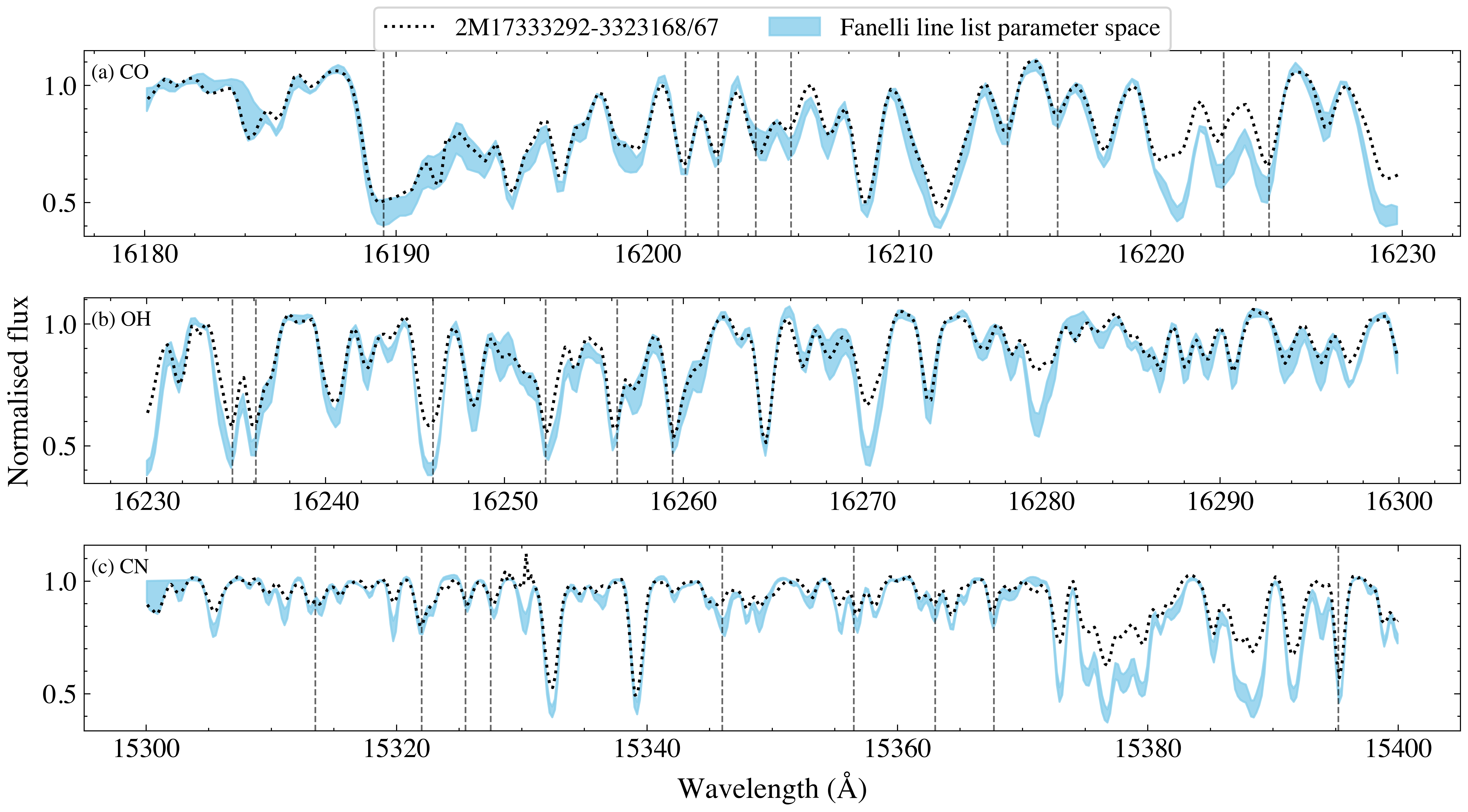}
    \caption{Combined APOGEE spectra for star 2M17333292-3323168 indicated with the black dotted line. The blue shaded region indicates the parameter space covered by the spectra calculated with the stellar parameters and abundances adopted from \citet{Fanelli2024}. To check the inference made from Figure \ref{fig:spec} is not a consequence of using the APOGEE line list and solar reference abundances from \citet{Grevesse2007}, we model the spectrum using the VALD3 atomic linelist and \citet{Magg2022} solar reference abundances. The sources from which the molecular lines are from are shown in Table \ref{tab:molecularlinelists}. These encompass the molecular line list data for CO, CN and OH from VALD3, APOGEE and B. Plez. All in all, the use of different line list and solar reference abundances here did not significantly impact the spectral fit of the OH lines, the middle panel, as they are still overfit with respect to the observed spectrum.}
    \label{fig:valdmaggf24} 
\end{figure*}

The resulting spectra are shown in Figure \ref{fig:valdmaggf24} for the same spectral region displayed in the middle panel of Figure \ref{fig:spec}. We again find that the strengths of OH lines are systematically overestimated, despite adopting alternative solar reference abundances and line list.
Whilst there are small differences in these newly calculated synthetic spectra, they do not alter our interpretation that our adopted combination of stellar parameters and elemental abundances is a better match to the APOGEE data (Figure \ref{fig:spec}) for the star in question.

\section{Analysis}
\label{sec:analysis}

In this section, we quantify the similarity of the abundance pattern of Liller~1 to the Galactic bulge field in order to test the hypothesis that the progenitor of Liller~1 is a major contributor of stellar mass to the Galactic bulge stellar populations. 

The abundance ratios in the analysis include the following elements: C, N, O, Mg, Si and Ca as they are reliably determined by ASPCAP and have abundance data for Liller~1. Further elements, such as aluminium and sodium are not included in this paper because ASPCAP did not converge for those elements for the Liller~1 sample. 

\subsection{Comparing Liller~1 and bulge chemical compositions}
\label{sec:Lillervsbulge}

\subsubsection{Removing candidate second generation stars from the Liller~1 sample}\label{sec:remove}

Before comparing the chemical composition of Liller~1 stars to those of the Galactic bulge population, it is important that the cluster stars with abundance patterns of so called ``second generation'' (2G) GC stars are removed from the sample in comparisons involving elements influenced by the multiple populations phenomenon \citep[]{Gratton2012}. Since 2G stars are quite rare in the field \citep[]{Martell2010, Martell2016, schiavon2017, Horta2021b}, their inclusion in the Liller~1 sample would lead to artificial differences between the cluster and the field for those elements. 

\begin{figure}[H]
    \includegraphics[width=0.8\columnwidth]{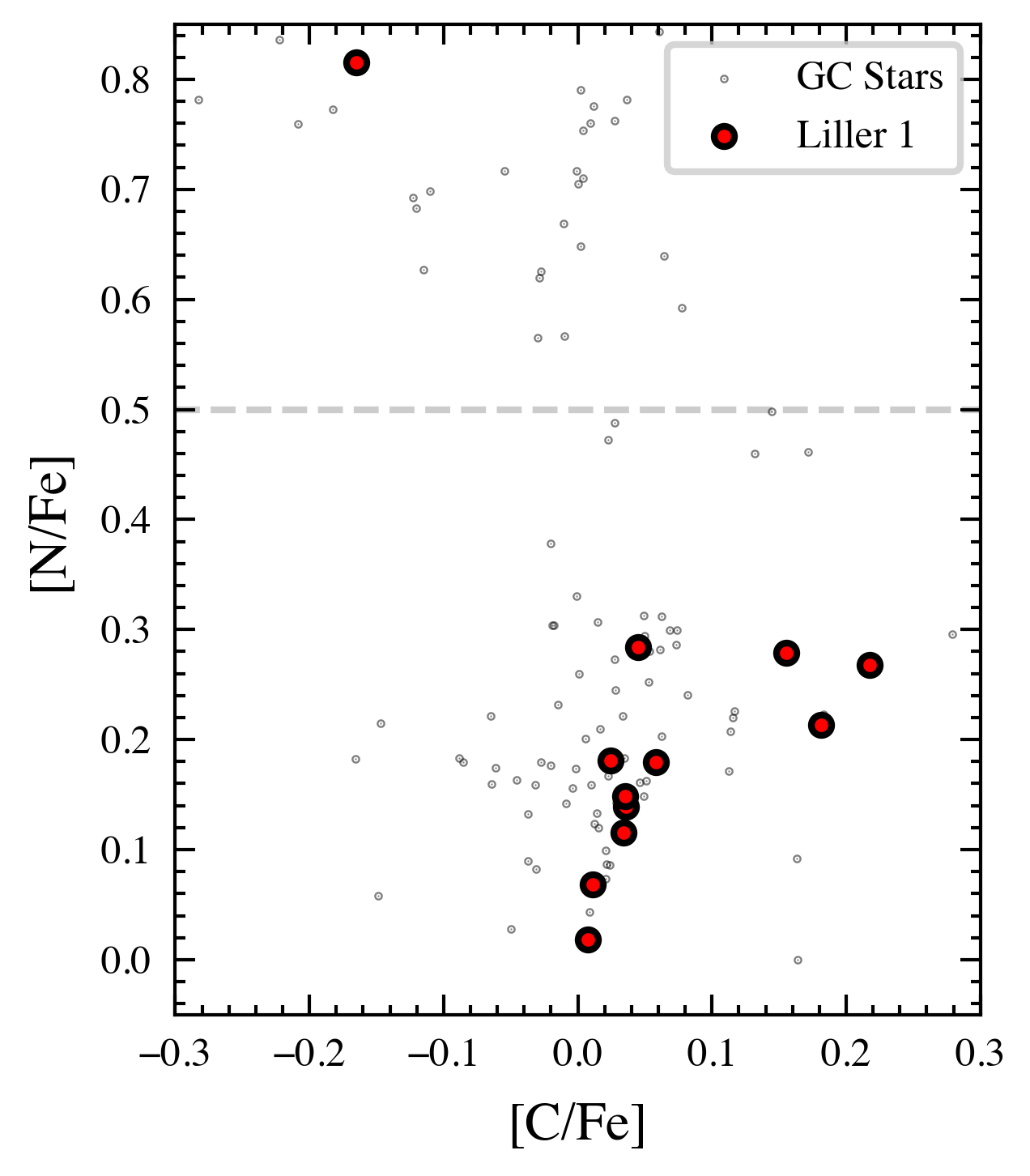}
    \caption{[N/Fe]-[C/Fe] of GCs from the VAC in the metallicity range ($\mathrm{[Fe/H]}$) spanned by Liller~1 stars in the sample. This includes GCS: NGC 6388, NGC 6582, NGC 6553, NGC 6441, Terzan 12, Palomar 10, Palomar 1. There is one Liller~1 star that exceeds the threshold $\mathrm{[N/Fe] = +0.5}$.}
    \label{fig:ncplane}
\end{figure}

For simplicity, we deem 2G stars in Liller~1 those for which $\mathrm{[N/Fe] > +0.5}$. Our Liller~1 sample is displayed on the [N/Fe]-[C/Fe] in Figure \ref{fig:ncplane}, where small gray symbols represent GC stars from the \citet{Schiavon2024} catalogue spanning the same [Fe/H] range covered by Liller~1 stars, which in turn are displayed as red circles. While the other GC stars display a standard distribution on the [N/Fe]-[C/Fe] plane with two sequences corresponding to 1G and 2G stars \citep[see, e.g.,][Fig.~7]{Schiavon2024}, our Liller~1 sample contains only one star exceeding this threshold (namely, 2M17332090-3322320). That object is removed from comparisons with the field populations involving elements C, N, O and Mg.

{  To confirm that this star is indeed N-rich and C-poor, we compare the observed spectrum to synthetic spectra generated using the method described in Section~\ref{sec:synthesis}. 
The base synthesis adopts the APOGEE-reported abundances and spectroscopic (as opposed to calibrated) stellar parameters. For comparison, we consider two cases. 
In the first, we adopt the C and N abundances typical of the other three metal-poor stars in Liller~1 that show field-like compositions. For this, we adopt [C/Fe] = +0.18 and [N/Fe] = +0.22. 
In the second, we test a hybrid case that combines the APOGEE C abundance with the lower value for the N abundance. Figure~\ref{fig:cnverify} shows the observed spectrum together with these model spectra. This exercise confirms that the APOGEE-reported abundances provide the best match to the observed spectrum across the three spectral regions. 
The other sets of  abundances yield spectra with systematically underestimated OH (first case) or CN (second case) lines.}

{  It is noteworthy that our sample of 14 Liller~1 stars contains only one star seemingly belonging to the 2G population.  Given its mass \citep[${\rm\sim2\times10^6~M_\odot,}$][]{saracino2015}, if Liller~1 was a regular globular cluster, one would expect that 2G stars would make up about 85\% of all its stellar content \citep[e.g.,][]{milone2017}.  
Such a large discrepancy is strongly suggestive that Liller~1 cannot be reasonably classified as a globular cluster.}

\begin{figure*}
    \includegraphics[width=1\textwidth]{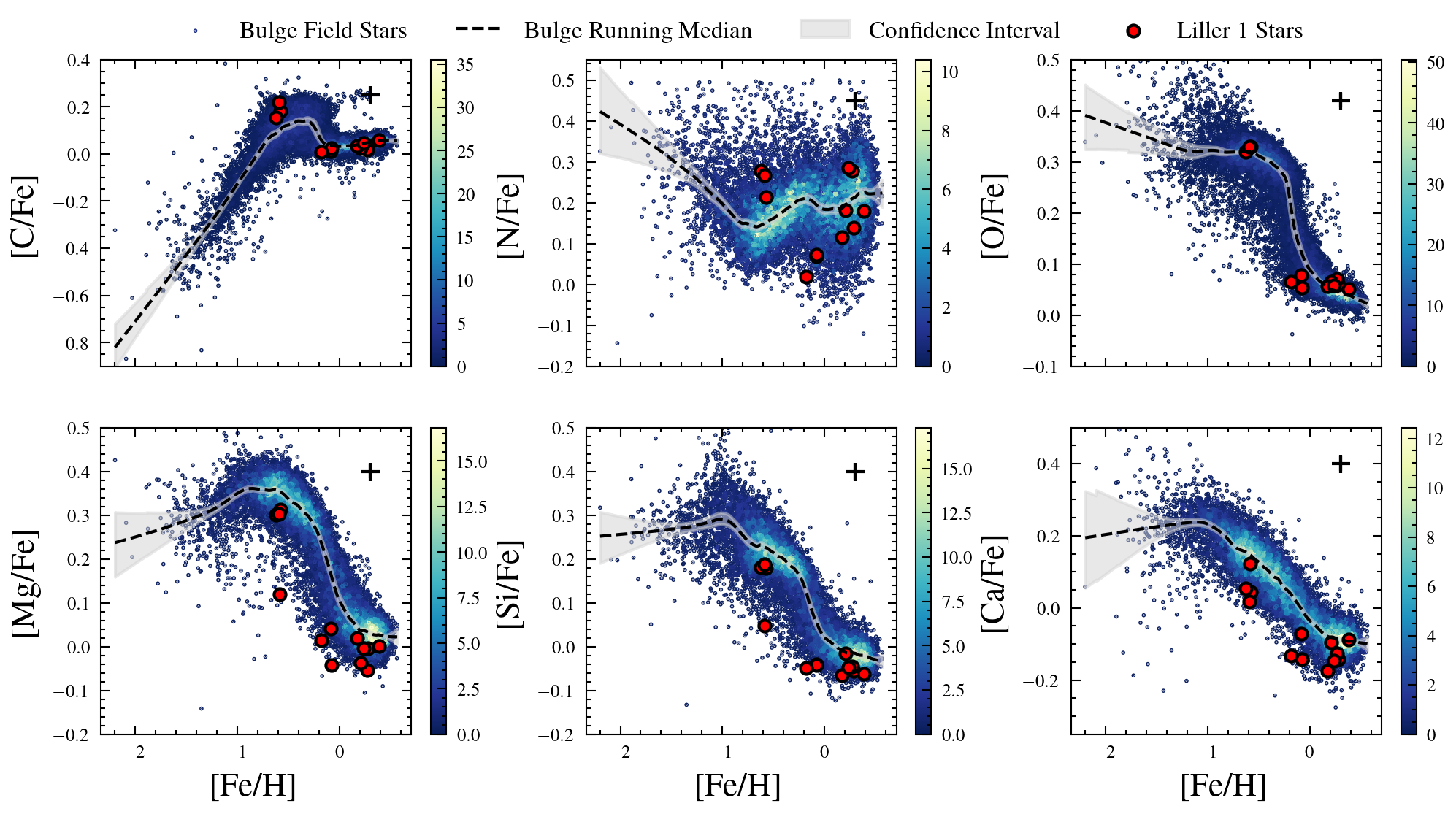}
    \caption{[X/Fe]–[Fe/H] scatter distribution of bulge field stars, within $\pm 0.25$ of the range of $\log{g}$ covered by the Liller~1 stars, colour-coded by local number density. The colour-coding method is described in the text. The sample of Liller~1 stars are shown in red. The running median of the bulge distribution is indicated with the black dashed line. The 95\% confidence interval on the running median is estimated by bootstrapping 25\% of the field stars to re-determine the running median. This is repeated 100 times to estimate the resulting spread for the confidence interval. The error in the median is shown in these plots for illustrative purposes. They are {\it not} used in the analysis (see text). The mean error on the abundances is shown in the top right of each panel. The data suggest that Liller~1 differs from the bulge systematically in Mg, Si, and Ca.}
    \label{fig:lowessfitsbulge}

\end{figure*}

\subsubsection{Qualitative comparison between Liller 1 and field stars}
\label{sec:qualify}

{  Figure \ref{fig:lowessfitsbulge} compares the bulge field population with Liller~1 stars in various [X/Fe]-[Fe/H] planes. The field data in these plots are displayed as follows. Initially, a 2D grid is defined on each chemical plane, consisting of ${\rm X,Y}$ bins. 
The number of data points within each bin is calculated, generating a 2D histogram. The latter, however, is not displayed, but rather each data point is colour coded by the data density corresponding to the bin within which it is located.  
In this way, both the 2D density information and the distribution of individual data points in low density regions is preserved (unlike the case of 2D histograms that miss the latter).
The field stars included in these plots are only those included within the $\log g$ range covered by the Liller~1 sample $\pm 0.25$ ($0.18 < \log g < 1.75$) for reasons which are detailed below.}

The Liller~1 sample is overlaid in red. The running median for the bulge population is derived using the \textsc{statsmodels} locally weighted scatterplot smoothing (LOWESS) algorithm \citep[]{cleveland1979}. The smoothing is applied with a weighting fraction of 0.07 of the surrounding data and iterated three times. To estimate the 95\% confidence interval of the LOWESS fit, we perform a bootstrap resampling of 25\% of the data 100 times, calculating the spread of the resulting fits, shown as the grey-shaded region. Visual inspection of Figure \ref{fig:lowessfitsbulge} indicates that there are differences between Liller~1 and the distribution of the bulge stars~-~in particular for $\alpha$-elements Mg, Si, and Ca. The metallicity distribution of Liller~1 in Figure \ref{fig:lowessfitsbulge} displays three clear peaks around $\mathrm{[Fe/H]} \approx -0.6$~dex, $\mathrm{[Fe/H]}\approx -0.15$~dex and a high metallicity group around $\mathrm{[Fe/H]} \approx +0.3$~dex, which is consistent with results from \cite{Fanelli2024} and \cite{2025Ferraro}. 

\subsubsection{Quantifying abundance differences between Liller~1 and the field}\label{sec:quantify}

In order to compare the abundance patterns in a statistically robust manner, we adopt the differential method from \citet{taylor2022}. That approach quantifies the difference between the abundance patterns of Liller~1 and bulge stars, while minimising the impact of systematics in the ASPCAP abundance determinations, which are correlated with position along the red giant branch \citep[see, e.g.,][]{weinberg2022,eilers2022,Horta2023}. The method is based on the following statistic:

\begin{equation}
    \rho_X = \mathrm{median} \left( \frac {[X/\mathrm{Fe}]^{\mathrm{Lil1}}_i - [X/\mathrm{Fe}]^{\mathrm{Bulge}}_i}{\sqrt{\sigma^2[X/\mathrm{Fe}]^{\mathrm{Lil1}}_i + \sigma^2[X/\mathrm{Fe}]^{\mathrm{Bulge}}_i}} \right),
    \label{eq:rho_x}
\end{equation}

\bigskip
\noindent Where $[X/\mathrm{Fe}]^{\mathrm{Lil1}}_i $ and $\sigma^2 [X/\mathrm{Fe}]^{\mathrm{Lil1}}_i $ are the abundance ratio and the square of its error for element X in Liller~1 star $i$. $[X/\mathrm{Fe}]^{\mathrm{Bulge}}_i $ is the median of a bulge field subsample selected to resemble star $i$ in [Fe/H] and $\log g$ and $\sigma^2[X/\mathrm{Fe}]^{\mathrm{Bulge}}_i$ is the standard deviation about that median.

\begin{figure}
    \includegraphics[width=1\columnwidth]{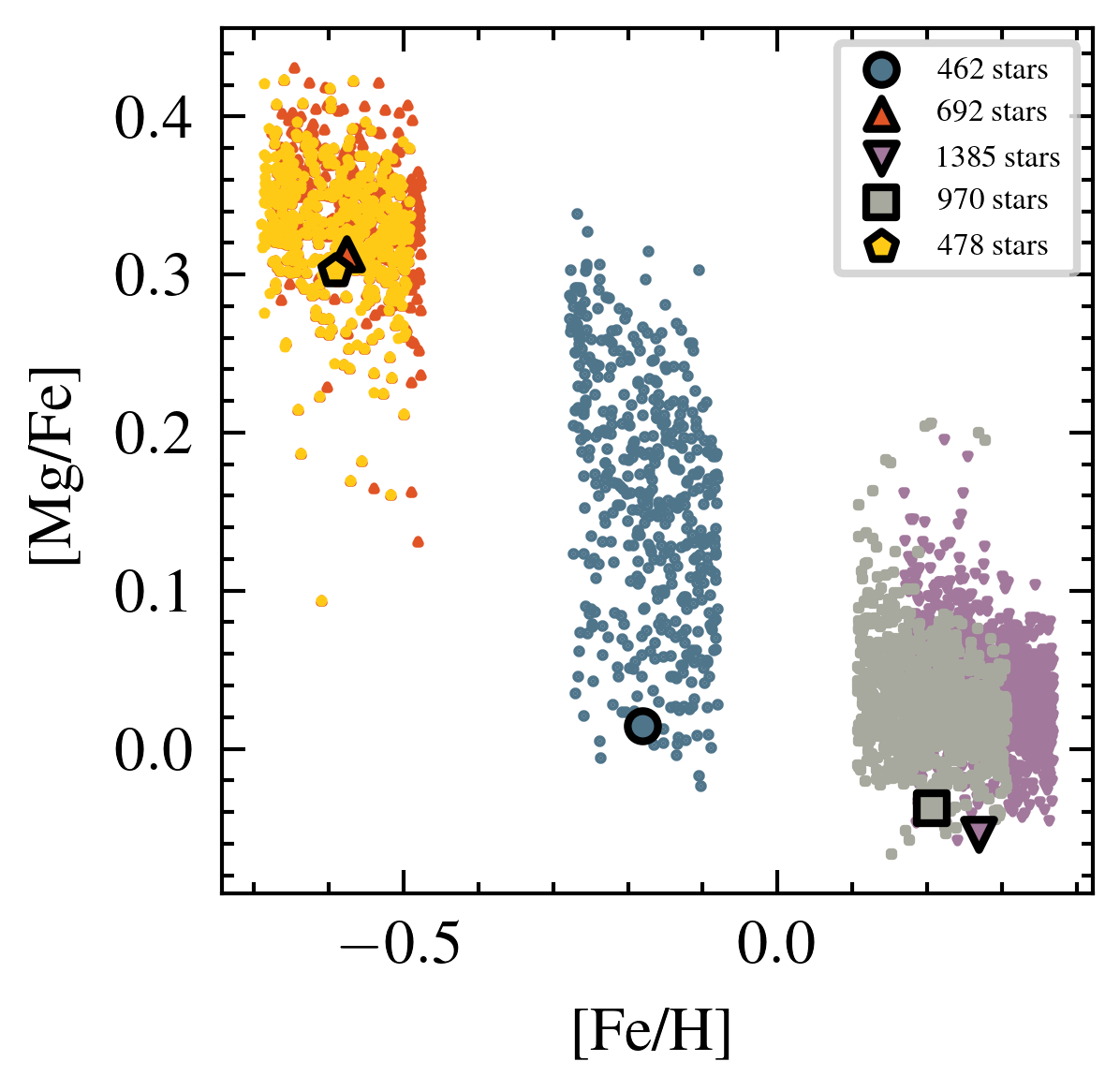}
    \caption{[Mg/Fe]-[Fe/H] distribution of 5 randomly selected Liller~1 stars and their corresponding stars in the field defined as similar stars in terms $\pm 0.1$~dex in [Fe/H] and $\pm 0.25$~dex in ${\log (g)}$ around each star $i$ in Liller~1. The groups of similar stars are colour-coded to overlap the colour of the corresponding Liller~1 star. This plot is to illustrate the comparison samples used in the analysis of the abundance offset between Liller~1 and the field. There are clear differences between Liller~1 stars and their similar counterparts in the field.}
    \label{fig:simstarsca}
\end{figure}

The bulge subsample selected for comparison with each star $i$ in Liller~1 is defined so as to contain objects differing from that star by no more than [Fe/H] $\pm 0.1$~dex and ${\log g}$ $\pm 0.25$~dex. Figure~\ref{fig:simstarsca} shows the Liller~1 stars along with the bulge subsamples adopted for the comparison with each of the cluster stars in the [Mg/Fe]-[Fe/H] plane for 5 of the Liller~1 stars. The reasoning for proceeding in this way is twofold: 

\begin{itemize}

\item First, as discussed by \citet{Weinberg2019}, systematic effects cause the stellar abundance determinations by the ASPCAP pipeline to show mild, but non-negligible correlations with surface gravity ($\log g$). For that reason, we compare each Liller~1 star with stars in the bulge field sample within the narrow range of the ${\log g}$ value ($\pm 0.25$~dex) to minimise spurious effects. 

\item Second, the surface gravity criterion in conjunction with the metallicity criterion ensures the abundance pattern of Liller~1 in this paper is being compared to the abundance pattern of similar stars from the Galactic bulge.

\end{itemize}

Following the method from \citet{taylor2022}, for each star $i$ in Liller~1, we calculate the offset between the median [X/Fe] of the similar stars and said star. This is weighted by the error on the abundance from APOGEE and the error on the median (Equation~\ref{eq:rho_x}). The $\rho_{\rm X}$ values for each abundance ratio for the Liller~1 sample are listed in Table \ref{tab:rhoxbulge}.

\begin{table}
\caption{$\rho_X $ values for the Liller~1 comparison with the bulge in units of standard deviations away from the median of the distribution of similar stars}
\resizebox{\columnwidth}{!}{%
\begin{tabular}{|l|c|c|c|c|c|c|}
\hline
[X/Fe] & C & N & O & Mg & Si & Ca\\
\hline
$\rho_X (\sigma)$ &-1.03 & 1.80 & 2.09 & -3.61 & -2.45 & -3.73\\
\hline
\end{tabular}%
}
\label{tab:rhoxbulge}
\end{table}

Next we quantify the significance of the $\rho_x$ parameter obtained for Liller~1. For that purpose we ask the following question: if Liller~1 were indeed a building block of the bulge, what are the odds that we extract a set of bulge stars with the chemical composition of Liller~1? 
More specifically, what are the chances that we would obtain a value of $\rho_{\rm X}$ comparable to that for Liller~1? To answer that question we extracted a large number of random bulge subsamples the same size as our Liller~1 sample, and calculated the value of $\rho_{\rm X}$ for each one of them. 
The subsamples were selected following a bootstrapping with replacement method. While the median value of $\rho_{\rm X}$ for the bootstrapped subsamples must by definition be very close to zero, the distribution of the $\rho_{\rm X}$ values for this bootstrap sample provides information on how significant the non-zero $\rho_{\rm X}$ values found for the Liller~1 sample actually are.
This is because, if the abundances of Liller~1 and the bulge distribution for element $\mathrm{X}$ deviate significantly, then the $\mathrm{\rho_X}$ value obtained for Liller~1 should differ from that of the medians of the random bulge samples in a statistically significant way. 

The number of similar stars in the bulge sample for each star in Liller~1 { (Figure~\ref{fig:simstarsca}}) ranges from 432 to 1385 stars. We implement the bootstrapping with replacement technique to produce 1000 realisations of 14 (13) stars randomly drawn from the 14 (13) subsamples of similar stars. 
The value of $\rho_{\rm X}$ is then calculated. For each star in the 1000 Liller~1-sized bulge random samples, we calculate the offset in the abundance from the same median and error on the median as used in the $\rho_{\rm X}$ calculation for the Liller~1 sample. 

\begin{figure*}
    \includegraphics[width=1\textwidth]{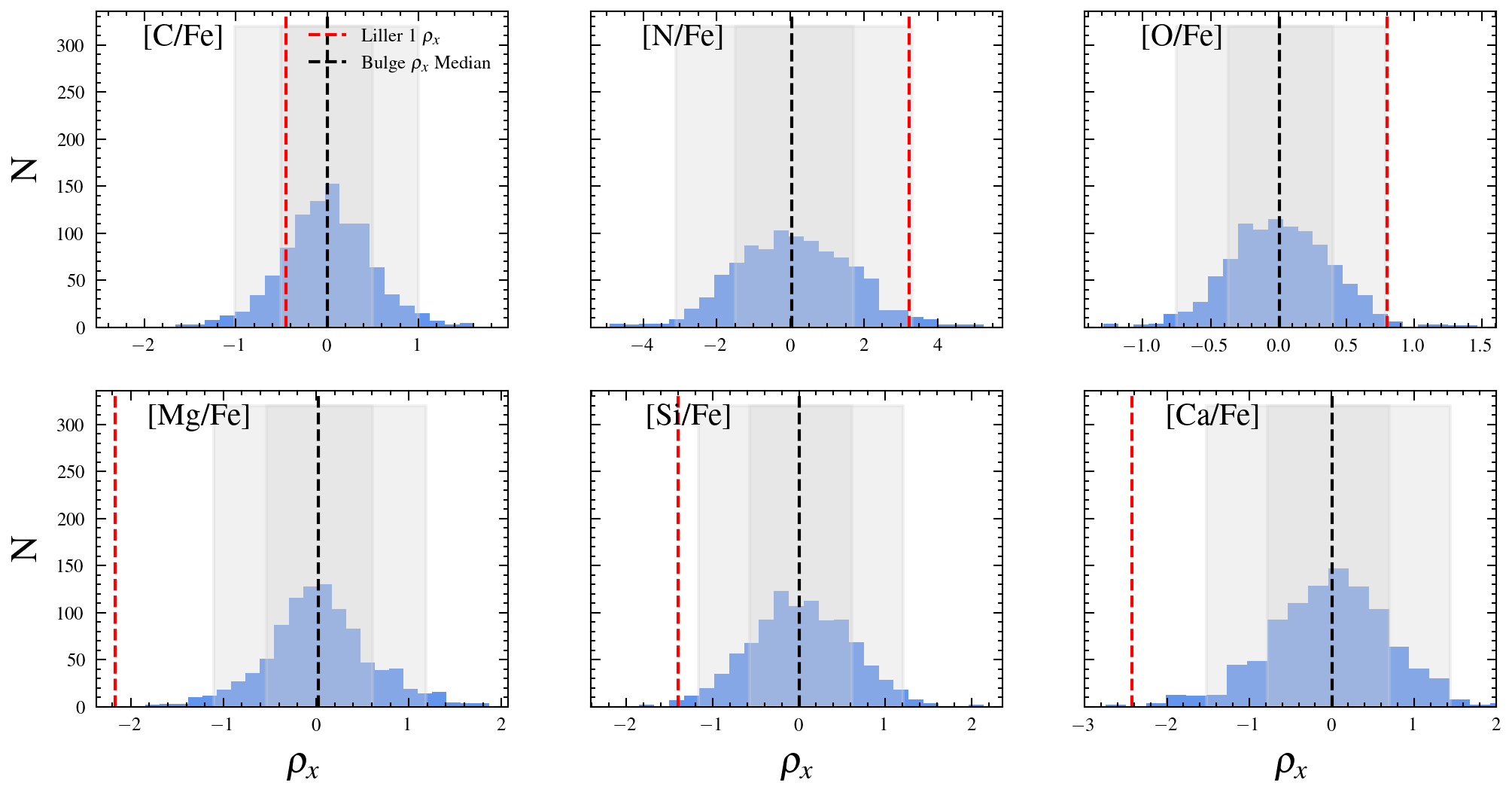}
    \caption{ The $\rho_X$ distributions of the bulge field similar stars in terms of $\pm 0.1$~dex and $\pm 0.25$~dex in [Fe/H] and ${\log g}$ around each star $i$ in Liller~1 is shown by the histogram, where the black dashed line indicates median of the $\rho_X$ distribution. The $\rho_X$ value obtained for Liller~1 is indicated with the red dashed line. The dark and light grey shading indicate the $1\sigma$ and $2\sigma$ intervals of the bulge field $\rho_X$ distribution.}
    \label{fig:rhobulge}
\end{figure*}

Figure~\ref{fig:rhobulge} shows the distribution of the $\mathrm{\rho_X}$ values obtained for Liller~1 and the 1000 random samples drawn from the bulge field. The median of the $\mathrm{\rho_X}$ distribution for the random samples ($\approx 0$, as expected) is marked by a black dashed line, while the red dashed line indicates the $\mathrm{\rho_X}$ obtained for Liller~1. The grey shadings indicate the regions within 1$\sigma$ and 2$\sigma$ of the median of the $\rho_X$ distributions for the bulge sample. 
While Figure~\ref{fig:rhobulge} shows that the abundance pattern of C, N and O deviates at the 1$\sigma$ level from the median of the field sample, the alpha elements Si, Mg and Ca show the most significant offset with them all being depressed in Liller~1 with respect to the Galactic bulge sample. Si differs at the 2$\sigma$ level and both Mg and Ca deviate from the bulge distribution of similar stars at the 3$\sigma$ level. 

From this we conclude that our data are consistent with Liller~1 and the Galactic bulge stellar populations being chemically distinct in a statistically significant way. The results of the analysis show that the progenitor of Liller~1 could not have been a {\it major} contributor of stellar mass to the Galactic bulge field as the differences in the abundance patterns indicate that its impact on the mean bulge chemistry, if any, was negligible.


{  
For completeness, we note that the same method was applied to abundances from DR16 and those from DR17 that were obtained by optimization of the APOGEE spectra against a \textsc{turbospectrum} spectral library.  
In the case of DR16, the statistical significance of the results was somewhat weakened, with only Mg differing between Liller~1 and the bulge by more than 2$\sigma$. 
However, in DR16, the sample of Liller~1 stars with abundances was reduced to a sample of 6-7 stars, with those differing from the bulge abundances the more strongly being absent from the catalogue.  The DR17-\textsc{turbospectrum} analysis was based on a library calculated adopting the spherically symmetrical MARCS model atmospheres. 
Visual inspection of the distribution of bulge field stars on the various chemical planes suggested the presence of seemingly unphysical substructure, particularly in the metal rich regime of the field. 
In addition, the APOGEE team recommends use of the DR17's \textsc{synspec}-based spectral library, with NLTE corrections for some of the key elements in our analysis (Mg and Ca), so we have good reason to take the results based on DR17 data as the most reliable.}


{  
\subsubsection{Robustness tests}
\label{sec:robust}
}

{  We test the robustness of the criteria for defining similar stars by, (i) adopting a more stringent sample with ${\log g}\pm 0.1$~dex and (ii) simultaneously introducing an effective temperature criterion of $\rm{T_{eff}} \pm 150~K$. 
One of the Liller~1 stars did not have a sample of similar stars within the specified temperature range and was removed from the sample. The statistical comparison for both tests yielded results consistent with the original analysis. The largest difference was a $\sim+0.3~\sigma$ offset in N and O. Thus, we retain the original comparison criteria in order to simultaneously maximise the number of similar stars within the comparison field whilst maintaining the reasoning discussed above.}

{  Finally, as pointed out in Section~\ref{sec:candidates}, the abundance uncertainties reported by APOGEE are exceedingly small. They are in fact measurements of the precision of APOGEE abundances obtained from very high S/N spectra. 
While internal systematics may still be present, such as $\log g$-dependent trends, or zero-point offsets due to input physics employed in the abundance analysis, they are mitigated by restricting the comparison to stars of similar $\log g$ and by the overall extreme homogeneity of the very high S/N data.
Nevertheless, to test the robustness of our $\rm{\rho_x}$ statistic to abundance error assumptions, we repeated the analysis by randomly assigning larger abundance uncertainties between 0.08 and 0.12~dex for each Liller~1 star. This adjustment did not lead to a change in the final offset values by more than 0.2~$\sigma$ for any of the elements.}

\bigskip

\subsection{A few clarifications on the method}
\label{sec:clarify}

A recent study \citep[]{origlia2025} has questioned the method employed in our quantitative analysis, presented originally by \citet{taylor2022}, where they compared the chemistry of Terzan~5 with that of the bulge field. In view of that, we take this opportunity to offer a few clarifications regarding the parameters used in Equation~\ref{eq:rho_x} and the method in general. 

We first clarify that the running median displayed in Figure~\ref{fig:lowessfitsbulge} is {\it not} used to asses the similarity in abundance pattern between Liller~1/Terzan~5 and the field. Instead, the $[X/\mathrm{Fe}]^{\mathrm{Bulge}}_i$ term is the median of the values for bulge stars with similar [Fe/H] and $\log g$ (Figure~\ref{fig:simstarsca} again) to each of the abundances ($[X/\mathrm{Fe}]^{\mathrm{Lil1}}_i$) of individual members of Liller~1.
By proceeding in this way, we guarantee that the comparison is performed between stars belonging to similar stellar populations (i.e., similar metallicity, since the effect of age on the giant branch morphology is negligible for our purposes) and at similar evolutionary stage. 

On a similar note, it is important that we elucidate our use of the 95\% confidence interval around the median, displayed as the shaded gray area in all panels of Figure~\ref{fig:lowessfitsbulge} \citep[Figure~3 in][]{taylor2022}. That confidence interval is {\it not} used as an estimate of the scatter in the distribution of the bulge abundances at fixed [Fe/H] (the $\sigma^2[X/\mathrm{Fe}]^{\mathrm{Bulge}}_i$ term in Equation~\ref{eq:rho_x}). 
It is plotted with the sole purpose of illustrating how well the running median is determined. Instead, $\sigma^2[X/\mathrm{Fe}]^{\mathrm{Bulge}}_i$ is obtained from the subsample of bulge stars picked for comparison with each Liller~1 star, displayed in Figure~\ref{fig:simstarsca}. 

Finally, we stress that we resort to the \citet{taylor2022} method because our sample is relatively small (though not as small as Taylor et al.'s sample of Terzan~5 members). Therefore, we need a reliable method that enables us to quantify the chemical composition difference between Liller~1 and the bulge population. 
In addition, the method provides an assessment of the significance of the difference measured, by assessing the likelihood of drawing a Liller~1/Terzan~5-like sample from the parent sample of all bulge stars. This method thus delivers a robust statistical result, which strongly suggests that neither Terzan~5 nor Liller~1 is a {\it major} contributor to the stellar make up of the Galactic bulge.

\subsection{Comparison with the Galactic disc}
\label{sec:diskresults}

In this Section, we test a scenario according to which Liller~1 is part of a massive clump that formed in the Galactic disc at high redshift and later migrated inwards. Our hypothesis is that if the progenitor of Liller~1 originated from the Galactic disc then chemical similarities between the disc field and the Liller~1 candidates are expected. The procedure above is thus repeated for the disc field populations as defined in Section~\ref{sec:disksample}.

To ensure a robust comparison, we divide the populations of Liller~1 and the disc into high and low-$\alpha$ populations. We use the criteria established by \citet{Weinberg2019} to distinguish between the high--$\alpha$ and low-$\alpha$ disc populations, defined as follows:

\[
\left\{
\begin{aligned}
    \mathrm{[Mg/Fe]} &> 0.12 - 0.13 \times \mathrm{[Fe/H]}, \quad \mathrm{[Fe/H]} < 0 \\
    \mathrm{[Mg/Fe]} &> 0.12, \quad \mathrm{[Fe/H]} > 0
\end{aligned}
\right.
\]

\begin{figure*}
    \includegraphics[width=1\textwidth]{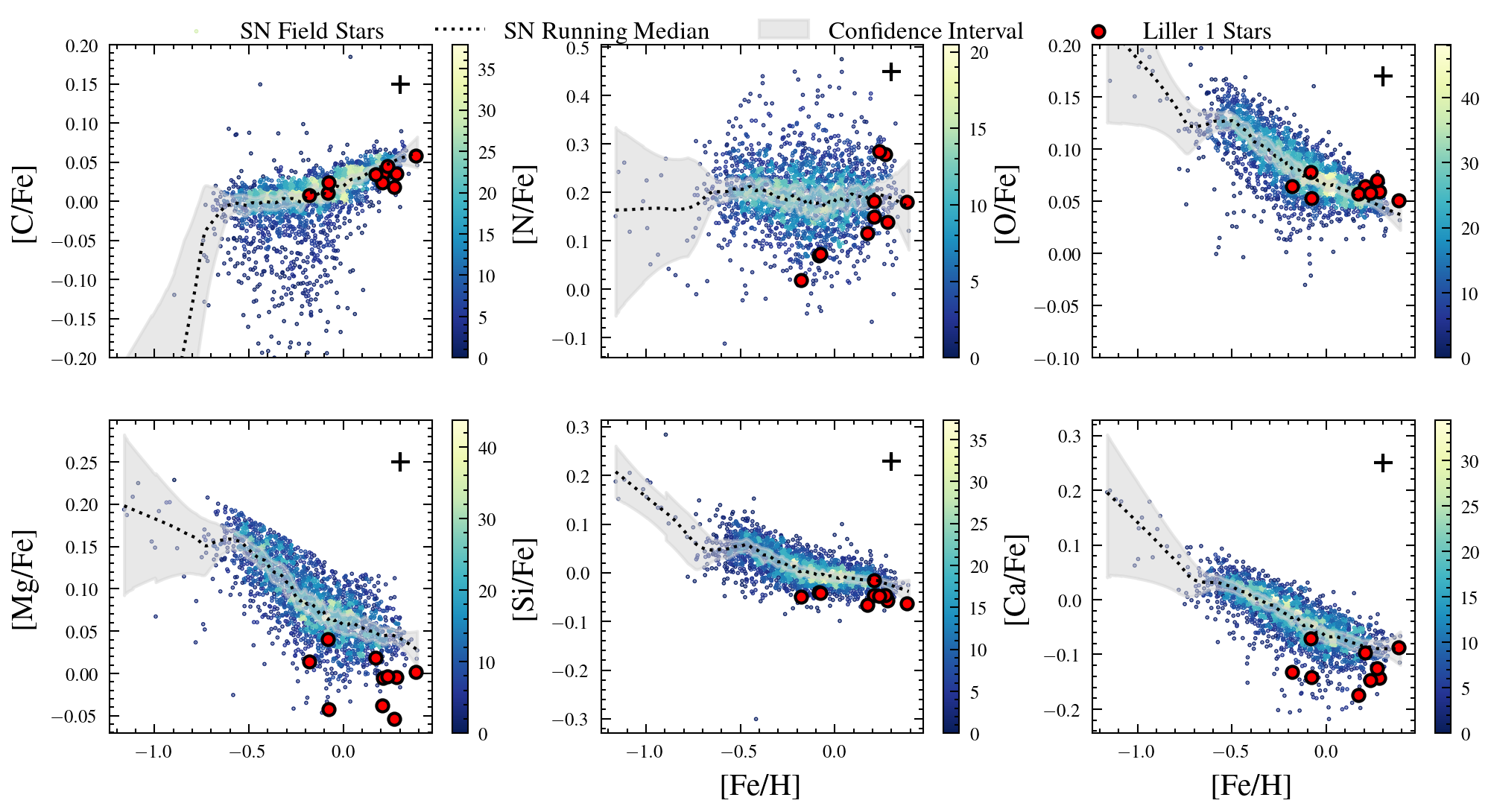}
    \caption{{  The [X/Fe]-[Fe/H] scatter plot with the number density of solar neighbourhood field stars described by the colour bar and Liller~1 stars shown in red}. The respective Solar neighbourhood running median is shown with the black dashed line.}
    \label{fig:lowesssn}
\end{figure*}

However, we do not include a comparison for the high-$\alpha$ stars in Liller~1 with the high-$\alpha$ disk, as only three stars in our Liller~1 sample meet this criterion (four for Ca and Si with the inclusion of the N rich star), which renders a robust analysis impractical. The ten remaining Liller~1 candidates are contrasted with the Solar neighbourhood low $\mathrm{\alpha}$ sample in the chemical planes in Figure \ref{fig:lowesssn}. The ten Liller~1 members are adopted for the comparisons in C, N, O, Si, and Mg and nine members in the case of Ca. In these plots the 2D histogram indicates the Solar neighbourhood sample in the narrow range $\log g \pm 0.25~\mathrm{dex}$ around the mean of Liller~1.

Visual inspection of the abundance planes indicates that there are differences in the elemental abundances between Liller~1 and the Solar neighbourhood field, with Ca, Mg and Si showing the most offset in the abundance. To quantitatively compare the chemical compositions of Liller~1 and the Solar neighbourhood, the $\rho_X$ statistic is again adopted to assess the chemical differences.

\begin{figure*}
    \includegraphics[width=1\textwidth]{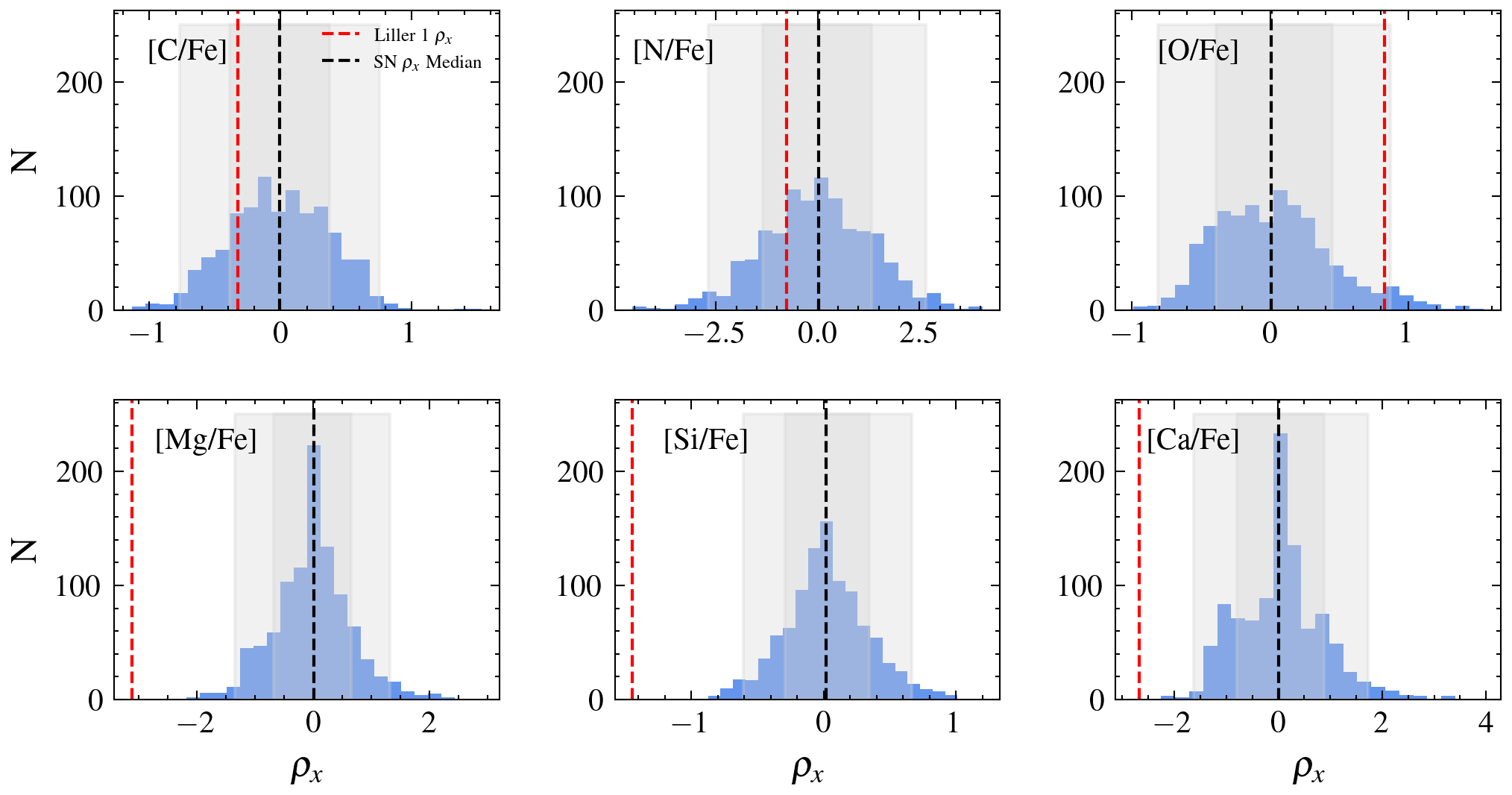}
    \caption{Histograms of the $\rho_X$ distributions of the Solar neighbourhood random samples. The median of the randomly sampled distributed is indicated with the black dashed line. The light and dark grey shading indicate the $1\sigma$ and $2\sigma$ intervals respectively. The Liller~1 $\rho_X$ is indicated with the red dashed line.}
    \label{fig:rhosn}
\end{figure*}

The $\rho_X$ distributions for the Solar neighbourhood are shown in Figure \ref{fig:rhosn} with the running median as a function of [Fe/H] of the field sample indicated by the black dashed line. Liller~1 differs from the stellar population of the Solar neighbourhood by at least 2$\sigma$ for O and 3$\sigma$ for Ca. Mg and Si differ at the 4$\sigma$ level with C and N showing no significant difference chemically to the Solar neighbourhood.

We further compare the abundance pattern of Liller~1 to the inner and outer low-$\alpha$ disc field samples as defined in Section \ref{sec:disksample}. The results are shown in Appendix~\ref{sec:Appendix}. The ten Liller~1 candidates are contrasted with the inner disc low $\mathrm{\alpha}$ sample in the chemical planes in Figure~\ref{fig:lowessid} with the $\rho_X$ distributions for the inner disc fields shown in Figure~\ref{fig:idrho}. The abundance pattern of Liller~1 deviates from the inner disc field at the 2$\sigma$ level for Mg, Si, Ca and the 3$\sigma$ level for O. 

The ten Liller~1 candidates are contrasted with the outer disc low $\mathrm{\alpha}$ sample in the various chemical planes in Figure \ref{fig:lowessod}. We do not proceed with the $\rho_X$ comparison between Liller~1 and the outer disc population, as most of Liller~1 stars in this comparison have a limited sample of disc stars within the constraints of [Fe/H] ± 0.1~dex and $\log g$ ± 0.25 for each Liller~1 star $i$. Indeed, the number of stars similar to Liller~1 stars in the outer disc sample varies from 1 to 1168, depending on the $\mathrm{[Fe/H]}$, with the higher metallicity stars having the lowest number of similar stars within the disc population. This fact alone argues for an important chemical difference between Liller~1 and the outer disc population, which does not reach metallicities as high as those of Liller~1 most metal-rich members. 

\begin{figure}
    \includegraphics[width=1\columnwidth]{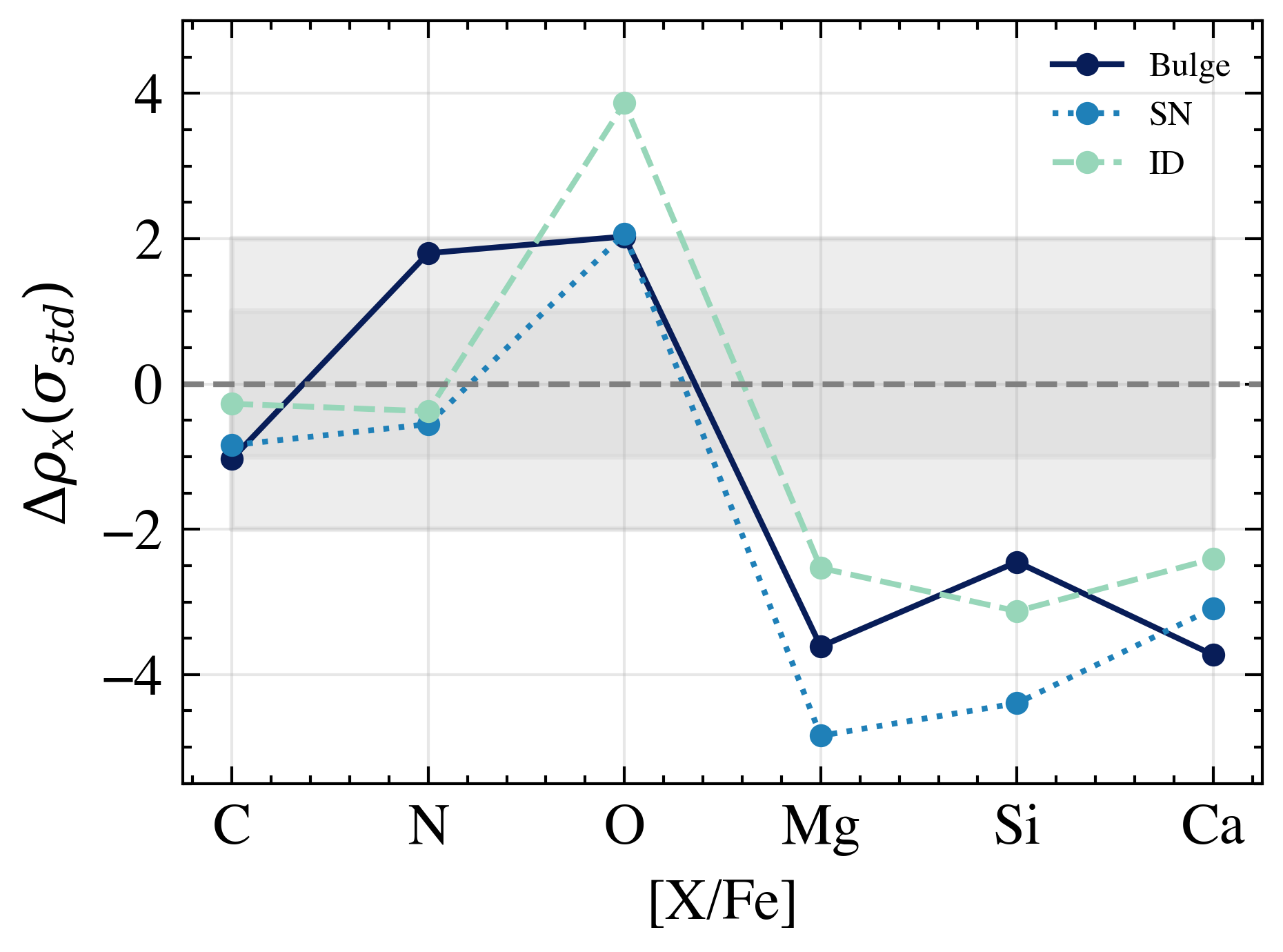}
    \caption{Results of the random sampling technique shown in separation $\Delta \rho_X$, in units of standard deviations, between the median of the random sampled bulge, Solar neighbourhood (SN) and inner disc (ID) $\rho_X$ distribution and the Liller~1 $\rho_X$ for each element. The gray shading indicates the region of the $\rho_X$ distribution that is one and two standard deviations away from the $\rho_X$ median obtained from the field.}
    \label{fig:results}
\end{figure}

The results of these comparisons (minus that for the outer disk) are summarised in Figure \ref{fig:results}, where the deviation of $\rho_{\rm X}$ statistic for the bulge, solar neighbourhood and inner disk is plotted in units of standard deviation for the various elements considered. We find that Liller~1 has lower [Mg/Fe], [Si/Fe], and [Ca/Fe] than all the populations considered. Overall, Liller~1 displays lower [$\alpha$/Fe] than the high- and low-$\alpha$ disk at all metallicities. Abundance patterns such as this result typically from star formation histories of dwarf galaxies, characterised by low star formation efficiency \citep[see, e.g.,][]{Tolstoy2009, Mackereth2019, Hasselquist2021, fernandes2023, Mason2024}.

In conclusion, our results indicate that the abundance pattern of Liller~1 differs significantly, in a statistical sense, from that of the Galactic disc samples within this study; this suggests that the progenitor of Liller~1 is unlikely to have originated from the Galactic disk. 

\section{The Nature of Liller~1}
\label{sec:natureofliller1}
In this section we discuss the implications of our results for scenarios proposed in the literature to explain the properties of Liller~1. The hypotheses we focus on are: Liller~1 as a rejuvenated GC, Liller~1 as an accreted system and Liller~1 as a disc instability clump.

\subsection{Liller~1 as a disc-instability clump}
\label{sec:diskinsta}

Initially classified as bulge GCs, Liller~1 and Terzan~5 both host stellar populations with a spread in age and metallicity. Furthermore, in both systems the young stellar populations are centrally segregated within the clusters; given the high mass of the systems, this was proposed to be the results of a self enrichment scenario. 
From these characteristics, \citet{ferraro2016, ferraro2021} identified both stellar systems as primordial building blocks of the bulge of the MW. It is important that we emphasise, however, that although the difference in chemical compositions between Liller~1 and the bulge field indicates that Liller~1 was not a {\it major} contributor of stellar mass, it does not rule out the progenitor of Liller~1 being a {\it minor} building block of the bulge. 
By assuming an initial mass of $\mathrm{10^9 M_\odot}$, \citet{ferraro2021} proposed that as many as ten such systems may have contributed to the total stellar mass content of the bulge. In this scenario, a difference in chemical composition between Liller~1 and the bulge is feasible assuming Liller~1 only contributed 1/10th of the total stellar mass content of the bulge.

However, \citet{taylor2022} have recently shown the chemical composition of Terzan~5 and the bulge differ at statistically significant levels (see discussion in Section~\ref{sec:clarify}). These findings indicate that, under the bulge fossil fragment scenario, the other remaining building bocks of the bulge would have undergone different star formation histories to compensate for the observed differences in chemistry between the field population, Liller~1 and Terzan~5. 
These systems hypothesised to be building blocks of the bulge are proposed to be the remnants of massive clumps, formed at high redshift from in situ disc instabilities that migrated inwards due to dynamical friction, and finally coalesced with other such systems to form the bulge \citep[]{noguchi1998, Ceverino2010, Debattista2023}.

Cosmological simulations predict clump masses in the range of $\mathrm{10^8 - 10^9~M_\odot}$ \citep[]{Inoue2017}; simulations further show that a fraction of the clumps can be long lived and survived feedback during their inward migration with some clumps having ex situ origins \citep[]{Mandelker2014, Mandelker2017}. 
Moreover, \citet{Debattista2023} have shown that clumps developed by hydrodynamical simulations provide a way of reproducing the abundance pattern displayed by bulge field stars through a combination of in-situ stars with stars from migrated clumps. 
Most importantly, simulations by the same group, \citep[]{Clarke2019} predict that such clumps are characterised by very high star formation rate density, giving birth to stars with high [$\alpha$/Fe], indeed higher than that of the low-$\alpha$ sequence. This prediction is at odds with our results from Section \ref{sec:Lillervsbulge}, which shows that Liller~1 stars have {\it lower} [$\alpha$/Fe] than both the bulge and disk, as measured by Mg, Si, and Ca abundances.

From the observational side, clumps observed in high-redshift progenitors of Milky Way-like galaxies are found to have average masses of $\sim 10^8 \mathrm{M_\odot}$ \citep[]{Livermore2012, Livermore2015, 2018Cava}. \citet{huertas2020} estimated the stellar masses of 3,000 clumps in galaxies within the CANDELS survey \citep[]{Grogin_2011, Koekemoer2011}, reporting masses in the range $\mathrm{10^{7-9}~M_\odot}$. More recently, \citet{Claeyssens2023} extended this, identifying clumps with masses spanning $\mathrm{10^5 - 10^9 M_\odot}$.

It is evident from our statistical analysis that the abundance pattern of Liller~1 is significantly different from those of all the disc regions examined in this study. This difference in chemistry could potentially be explained by the existence of a chemical composition gradient within the progenitor of Liller~1. In Section~\ref{sec:darkm} we discuss that hypothesis, arguing that such a chemical composition gradient is unlikely. In conclusion, our results seem to be inconsistent with the scenario according to which the progenitor of Liller~1 was a disc instability clump.

\subsection{Liller~1 as a rejuvenated globular cluster}
\cite{Mckenzie2018} and \cite{bastian2022} proposed that iron and age complex clusters may be the result of GC orbits crossing paths with molecular clouds. Massive clusters in the inner Galaxy may have orbits that align with giant molecular clouds, leading to gas accretion. If the accretion is rapid or dense enough, it can overcome feedback from the original stellar population and form a new generation of stars. This is possible only providing the cluster is able to survive disruption.

\citet{Mckenzie2018} found the probability of such an interaction to be highest for massive GCs with disc like orbits as found for Liller~1 \citep[]{Baumgardt2019}. Under this scenario, Liller~1 would have undergone largely discrete star formation events. The best fit star formation history for Liller~1 is in contrast with this scenario, as it consists  of a main peak, with underlying low levels of star formation followed by two additional bursts \citep[]{Dalessandro2022}.  

In addition, the cluster rejuvenation scenario predicts that the chemical composition of populations of stars formed from the accreted gas would reflect that of the Galactic disk, with the original stellar population reflecting the composition of the GC at the time of formation. Unfortunately, our sample is not large enough to test quantitatively the similarities between subsamples of Liller~1 stars at different metallicities and their field counterparts. Nevertheless, a qualitative statement can be made as follows. This scenario would result in similarities between Liller~1 stars and the field to be greatest on the high metallicity end, which contains stars formed from this putative disk gas. However, inspecting Figures \ref{fig:lowesssn}, \ref{fig:lowessid}, \ref{fig:lowessod} we can see that the high metallicity regime is where the differences are the greatest in all Galactic fields investigated. All in all, our results seem to be inconsistent with that hypothesis.

\subsection{Liller~1 as the result of a globular cluster merger}
Another scenario proposed by \cite{Gavagnin2016} and \cite{Khoperskov2018} to explain the existence of GCs that contain a significant metallicity spread and several peaks in the metallicity distribution function is that these systems are a product of cluster mergers and interactions.
Using N-body simulations, \citet{Khoperskov2018} showed that one to two collisions per Gyr are expected for a GC system following a density distribution matching that of the MW disk, with the same scale length and height.
In this case, Liller~1 could represent a rare case of a cluster-cluster merger. Such a scenario can be tested with detailed studies of the stellar kinematics and internal structure of Liller~1.
For instance, one should expect that the properties of the merging clusters would have played an important role in defining the structure of Liller~1 in a predictable way \citep[]{Gavagnin2016, Khoperskov2018, Battisti2019}, whereby the structure is the result of the properties of both the merging clusters and how they merge 
For example, a merger between clusters of different initial densities should result in a centrally concentrated population in the cluster: the more compact progenitor should retain most of its mass after the merger producing the centrally concentrated population whereas the progenitor with the lower initial density would have contributed only about 10\% of its mass to the cluster \citep[]{Khoperskov2018, Battisti2019}.

In this scenario, the different stellar populations within Liller~1 would reflect those of the parent GCs. In that case, one would expect that the respective populations would exhibit the abundance anti-correlations characteristic of GCs, which is at odds with the results of this study and those of \citet{Alvarez2024}, who found no evidence for the existence of anti-correlations within Liller~1.  
{  In fact, as pointed out in Section~\ref{sec:remove}, the fraction of 2G stars in our Liller~1 sample is much smaller than expected given its mass, which at face value rules out a GC merger origin.} Nonetheless, to complicate matters, \cite{schiavon2017b} identified the presence of multiple populations in the other bulge GC with characteristics similar to those of Liller~1, namely Terzan~5.

Therefore, we conclude that, while the cluster merger scenario can in principle explain some of the properties of Terzan~5, it is at odds with the existing data for Liller~1, both photometric and spectroscopic. Furthermore, there are no Galactic GCs with ages comparable to that of the youngest populations observed in both Liller~1 (1-3~Gyr) and Terzan 5 (4.5~Gyr), making it improbable that the properties of Liller~1 and Terzan 5 can be explained by the GC merger scenario \citep[]{2013VandenBerg}.

\subsection{The progenitor of Liller~1 as a dark matter dominated system}
\label{sec:darkm}
One further scenario invoked to explain the properties of Liller~1, suggests that the progenitor had an extragalactic origin. {  This is the one scenario that our data cannot easily rule out. However, it poses a number of difficult questions which we address in this section, admittedly in a speculative fashion.}  

In order to reproduce the chemical evolution of a system that hosts the distribution of both Liller~1 and Terzan~5 in the [$\mathrm{\alpha}$/Fe] plane, it is reasonable to assume that the system is massive and dark matter dominated. 
Under this assumption, following \citet{taylor2022} we utilise the progenitor mass estimates based on the EAGLE suite of simulations analysed by \citet{Mason2024}. The latter authors studied the correlation between the stellar mass of the progenitor and the [Fe/H] value corresponding to the position of the so-called ``$\mathrm{\alpha}$-knee'' in the ``$\mathrm{\alpha}$-Fe plane'', in the EAGLE suite of numerical simulations \citep[]{schaye2015,crain2015}. 
They showed that the simulations predict a correlation in good agreement with observations, while exhibiting significant scatter as the position of the $\mathrm{\alpha}$-knee is both dependent on the total mass and the details of the star formation history. From the distribution of Liller~1 stars in the [$\mathrm{\alpha}$/Fe] - [Fe/H] planes, we estimate the position of the $\mathrm{\alpha}$-knee to be within the range --0.6 to --0.2~dex. 
Adopting the slope and scatter of the \citet{Mason2024} relation, the stellar mass of the progenitor of Liller~1 is estimated to be within the range $\mathrm{10^8-10^{10} M_\odot}$. We next discuss the implications of this estimate on Liller~1’s contribution to the stellar mass content of the bulge, considering the two extremes of this mass range. 
\bigskip

{ 
\subsubsection{The case for an ${\rm M_\star \approx 10^{8}~M_\odot}$ progenitor}
\label{sec:lowmass}
}

We first consider the case whereby the mass of the Liller~1 progenitor was of the order of the lower limit of the above range. 
In that case Liller~1 would have been the NSC of an $\mathrm{M_\star\approx10^8~M_\odot}$ dwarf galaxy. 
This case is consistent with our results, since such a relatively low mass progenitor could have contributed its stellar mass content without significantly impacting the chemistry of the bulge. 
Furthermore, this mass prediction is in agreement with those proposed by other groups (see Section~\ref{sec:diskinsta}). 

{  In this context, we consider the {\it strawman} scenario recently proposed by \cite{mason2025} for the history of star formation and chemical enrichment of the well known $\omega$~Centauri ($\omega$~Cen) stellar system. It is proposed that $\omega$~Cen is the NSC of a dwarf galaxy long accreted to the Milky Way \citep{Zinnecker1988, Freeman1993}. The stellar population mix in $\omega$~Cen is explained by \cite{mason2025} as resulting from an initial burst of {\it in situ} star formation, followed by a very early gas-rich in-spiralling of metal poor globular clusters. The gas brought in by these clusters would have been responsible for the triggering of a second burst of star formation.  
The infalling gas, being originally enriched to the second-generation globular cluster abundance pattern\footnote{By processes hitherto unknown, see \cite{bastian_multiple_2018}}, gives birth to a stellar population with similar chemistry.  

Liller~1 has some of the features that fit this scenario. It hosts a metal-rich population with a low-$\alpha$ abundance pattern similar to those of dwarf galaxies. Moreover, our sample contains one star consistent with a 2G-like abundance. While foreground contamination and small-number statistics prevents a robust constraint on the fraction of 2G stars, their presence can in principle make Liller~1 compatible with the \cite{mason2025} NSC scenario.  

However, unlike $\omega$~Cen \citep{mason2025} and M54 \citep{Schiavon2024}, Liller~1 has a distinctive property: an absence of a metal-poor population, and in particular one displaying the multiple populations phenomenon. It is difficult to reconcile this result with the strawman scenario proposed by \cite{mason2025}.  
We speculate that Liller~1 could have been the NSC belonging to a galaxy that did not undergo an early spiralling in of metal-poor globular clusters. Moreover, Liller~1 would have been formed through the infall of fairly metal-rich gas. Although this scenario is somewhat contrived, it is nonetheless consistent with our data.}
\bigskip

{  
\subsubsection{The case for an ${\rm M_\star \approx 10^{9-10}~M_\odot}$ progenitor}
\label{sec:highmass}
}
Next we consider the case where the progenitor of Liller 1
was a $M_\star\approx\mathrm{10^9-10^{10} M_\odot}$ dwarf galaxy accreted to the MW. 
In this case, the system would have contributed significantly to the stellar mass budget of the Galaxy.  Moreover, given Liller~1's current orbital configuration, one could reasonably assume that a large fraction of that mass would end up in the inner Galaxy. 
In this case, it is expected that such a contribution to the mass of the bulge would result in similar abundance patterns between the two systems, in contrast to our results. 

The observed abundance differences could be accommodated if the progenitor galaxy hosted a chemical composition gradient where the rest of the stellar mass was enhanced in Mg, Si, and Ca relative to the nuclear star cluster. \citet{taylor2022} assessed the likelihood of such a scenario in the case of Terzan~5, by considering the chemistry of the existing nuclear cluster M54 relative to its host galaxy, the Sagittarius dwarf spheroidal (Sgr dSph). 
Both the core and tidal streams of the Sgr dSph have lower $\mathrm{[\alpha/Fe]}$ than the NSC \citep[][]{Hayes_2020, 2021Fernandez}, which goes in the opposite direction to what is needed in the case where Liller~1 is the NSC of an accreted dwarf galaxy \citep[see][for a discussion]{taylor2022}, were we to accommodate the discrepancy between its abundance pattern and that of the bulge. 
Therefore, the scenario according to which Liller~1 was the NSC of a dwarf galaxy accreted to the MW requires a star formation and chemical evolution history that is quite different than that of our best known NSC-dwarf galaxy pair. 

However, dwarf galaxies can display a range of possible behaviors with regards to [${\rm \alpha/Fe}$], including cases of negative, flat, or even positive [${\rm \alpha/Fe}$] gradients \citep[e.g.,][]{spolaor2010}. 
In the case where Liller~1 was the NSC of a high mass dwarf galaxy, the outer regions of Liller~1's putative progenitor were likely characterised by lower [Fe/H] than our sample of Liller~1 stars, thus contributing to the stellar content of the bulge in the more metal-poor regime (see \citet{Neumayer2020} and the references therein for a review).  {  Nevertheless, given all of the above objections to this scenario, we deem it quite unlikely.}

\bigskip

{ 
\subsubsection{Constraints from the age distribution of Liller~1 stars} \label{sec:ages}
}

{  Another important constraint on the problem comes from the stellar ages resulting from CMD studies of Liller~1. The age distribution from \cite{Dalessandro2022} seems to rule out a scenario invoking a relatively massive merger with the Milky Way to have happened about 1-3~Gyr ago, as there is no current evidence for such a recent merger.

Let us consider on the other hand that Liller~1 was the NSC of a low mass galaxy accreted to the Milky Way a long time ago. It might be the case that a young, super-solar metallicity population was formed after or during the merger event. 
In that case, it is conceivable that these younger stars could have been formed from a mix of gas from the Milky Way disk and Liller~1, which could be consistent with our data.

Therefore, it may be possible to accommodate the existing data with a scenario whereby Liller~1 was the NSC of a low mass dwarf accreted to the Milky Way a long time ago, provided that the particulars of Liller~1's host interaction history with the MW and the resulting star formation history may produce a super-solar metallicity population with low $\alpha$ abundances.
}

\bigskip

In conclusion, the scenario according to which Liller~1 was the NSC of a relatively low mass ($\approx 10^8 M_\odot$) progenitor is the only one that our (admittedly limited) data set does not rule out. It is critical that tests be devised to verify this proposition. 
Conceivable probes would consist of searches for a chemical association with known substructures \citep[e.g.,][]{horta_chemical_2023}, or a dynamical association with known streams \citep[e.g.,][]{ibata2024}. 
In addition, the feasibility of this scenario can only be understood after some chemodynamic modelling is available. At a minimum, one would need two assessments. On one hand, an estimate of the impact of such a minor merger on the kinematics of the disk. On the other a detailed model of the chemistry of the stellar population resulting from the putative star formation event.

\bigskip 
\bigskip 

\section{Conclusions}~\label{sec:conclusions} 
Liller~1 was originally classified as a globular cluster. However, photometric and spectroscopic observations of Liller~1 revealed that its stellar populations display a range of age \citep[1-3~Gyr and 12~Gyr;][]{ferraro2021} and metallicity \citep[$\mathrm{-0.48 \simless [Fe/H] \simless +0.26}$~dex;][]{crociati2023}. 
This complex nature, in conjunction with the central segregation of the younger population, means Liller~1 was identified as a potential building block of the Galactic bulge---a bulge fossil fragment \citep[]{ferraro2009, ferraro2021}. We used APOGEE data to perform a quantitative comparison of the chemical compositions of Liller~1 stars to those of the Galactic field populations. Our results put to test the hypothesis that Liller~1 has contributed relevantly to the stellar content of the Galactic bulge, as proposed by other groups. We also examine various hypotheses for the origin of Liller~1. We summarise our conclusions as follows:

\begin{itemize}
\item We conducted a test of the hypothesis that Liller~1 is a building block of the Galactic bulge, through a comparison of the abundance pattern of Liller~1 to that of the bulge field populations. 
The abundance data for the fourteen Liller~1 members came from the APOGEE-2 DR17 value-added catalogue of globular cluster members \citep[]{Schiavon2024}, whereas the data for the bulge and disk come from the APOGEE DR17 catalogue \citep[]{abdurrouf2022}. 
By adopting a random sampling technique for the quantitative statistical comparison, it is shown that the abundance pattern of Liller~1 is chemically distinct to that of the Galactic bulge {  to a high degree of confidence.}
This suggests that the progenitor of Liller~1 was unlikely to be a {\it major} contributor to the stellar mass content of the bulge.

\item In a similar vein, we tested the scenario according to which Liller~1 was formed as a massive disc instability clump at high redshift, that then migrated towards the inner Galaxy. By showing that the chemical composition of Liller~1 stars differs in a statistically significant fashion from those of disc stars over a range of Galactocentric distances, we deem this scenario unlikely.

\item We discuss the scenario proposed by \citet{bastian2022} according to which Liller~1 is an old GC that underwent multiple bursts of star formation as a result of the accretion of fresh gas from encounters with giant molecular clouds. A reasonable expectation in this case would be that the chemistry of the old population of Liller~1 reflect that of the GC at the time of formation, while the young population shares chemical properties with their field counterparts. A robust test of this ``rejuvenation'' scenario, requires a denser sampling of Liller~1 stars in metallicity and age space than afforded by our sample. Nonetheless, the available data suggests there is minimal similarity between the field populations and metal rich populations of Liller~1. 

\item \citet{Khoperskov2018} proposed systems such as Liller~1 and Terzan~5 are the result of GC~-~GC mergers and interactions. 
In this scenario, the different populations of stars would have the chemistry of their respective GCs at the time of formation. The difficulty with this scenario is that one would expect the chemical properties of the merging GCs, such as the multiple populations phenomenon, to be detectable in the resulting system. However, we find no indication of the presence of abundance anti-correlations characteristic of GCs in our data, at odds with that expectation. We note, however, that \cite{schiavon2017b} showed that the multiple populations phenomenon manifests itself in APOGEE data for Terzan~5. There clearly is a need for more data for these systems, over a broad range of metallicities, to resolve this apparent discrepancy. 

\item We speculate that Liller~1 could have been the nuclear star cluster of a satellite galaxy that merged with the Milky Way in the past. Following the analysis by \citet{taylor2022} and \cite{Mason2024} of the relationship between the [Fe/H] of the $\alpha$-knee and $\mathrm{M_*}$ in the EAGLE hydrodynamical cosmological simulations, the stellar mass of the progenitor would be in the range $\mathrm{10^8-10^{10}M_\odot}$. 
The disagreement between the chemistry of Liller~1 and bulge field populations rules out a progenitor on the high mass end of this range, unless Liller~1 presented an unusual positive radial gradient of [$\mathrm{\alpha}$/Fe].
On the other hand, assuming a progenitor mass on the small end of the acceptable range for the progenitor of Liller~1, $\mathrm{10^8M_\odot}$, the difference in chemistry between Liller~1 and the bulge can be understood. 
In this scenario, the progenitor of Liller~1 lost the majority of its stellar mass without making a significant contribution to the inner Galaxy. In short, we suggest that Liller~1 may have been a {\it minor} contributor to the stellar mass budget of the inner Galaxy, with a possible extra-Galactic origin. However, it is difficult to accommodate current knowledge of the chemical evolution of dwarf galaxies with the presence of super-solar metallicity stars in a $10^8 {\rm M_\odot}$ system. 

This hypothesis can be tested in at least two different ways, both aiming at identifying an association between Liller~1 and both existing substructures and those likely to be discovered in the future. On one hand, such searches can be based on statistical comparisons of chemical compositions such as the one conducted in this work. On the other, they can consist of searches for a dynamical association with streams or other phase space substructures.

\end{itemize}
Undeniably, future investigations of complex systems such as Liller~1 and Terzan~5 will benefit from larger samples of stars from the systems with detailed chemical abundances, PMs and RVs. The statistical comparisons in this study could then be refined with improved precision. In addition, we will be able to gain additional insights into these systems with a study comprised of a sample of Liller~1 stars spanning a wider range of abundances. On the theoretical side, it would be interesting to investigate the impact of a minor merger such as that implied by this hypothesis on the kinematics of the inner Galaxy and on the chemical evolution of the system. 

\section*{Acknowledgments}
The authors thank Katia Cunha for the guidance on the spectrum synthesis and Henrik J\"{o}nsson on the uncertainties of stellar parameters and elemental abundances in APOGEE data.  Bertrand Plez is thanked for making available his molecular line list in digital format. The authors also thank the anonymous referee for insightful comments and suggestions that improved the manuscript. S.K. gratefully acknowledges funding from UKRI through a Future Leaders Fellowship (grant no. MR/Y034147/1). J.G.F-T gratefully acknowledges the grants support provided by ANID Fondecyt Iniciaci\'on No. 11220340, ANID Fondecyt Postdoc No. 3230001, and from the Joint Committee ESO-Government of Chile under the agreement 2023 ORP 062/2023. D.M. acknowledges support from the Center for Astrophysics and Associated Technologies CATA by the ANID BASAL projects ACE210002 and FB210003, and by Fondecyt Project No. 1220724. D.M.N. was partially supported by the University of Iowa’s Year 2 P3 Strategic Initiatives Program through funding received for the project entitled “High Impact Hiring Initiative (HIHI): A Program to Strategically Recruit and Retain Talented Faculty.

\section*{Data Availability}
This work is entirely based on the 17th data release of the SDSS IV/APOGEE-2 survey and the APOGEE Value Added Catalogue of Galactic Globular Cluster Stars. All data are publicly available at: \nolinkurl{https://www.sdss.org/dr17/} and \nolinkurl{ https://www.sdss.org/dr18/data_access/value-added-catalogs/?vac_id=105}.

\bibliographystyle{mnras}
\bibliography{paper} 

\appendix
\section{Some extra material}
\begin{figure*}
    \includegraphics[width=1\textwidth]{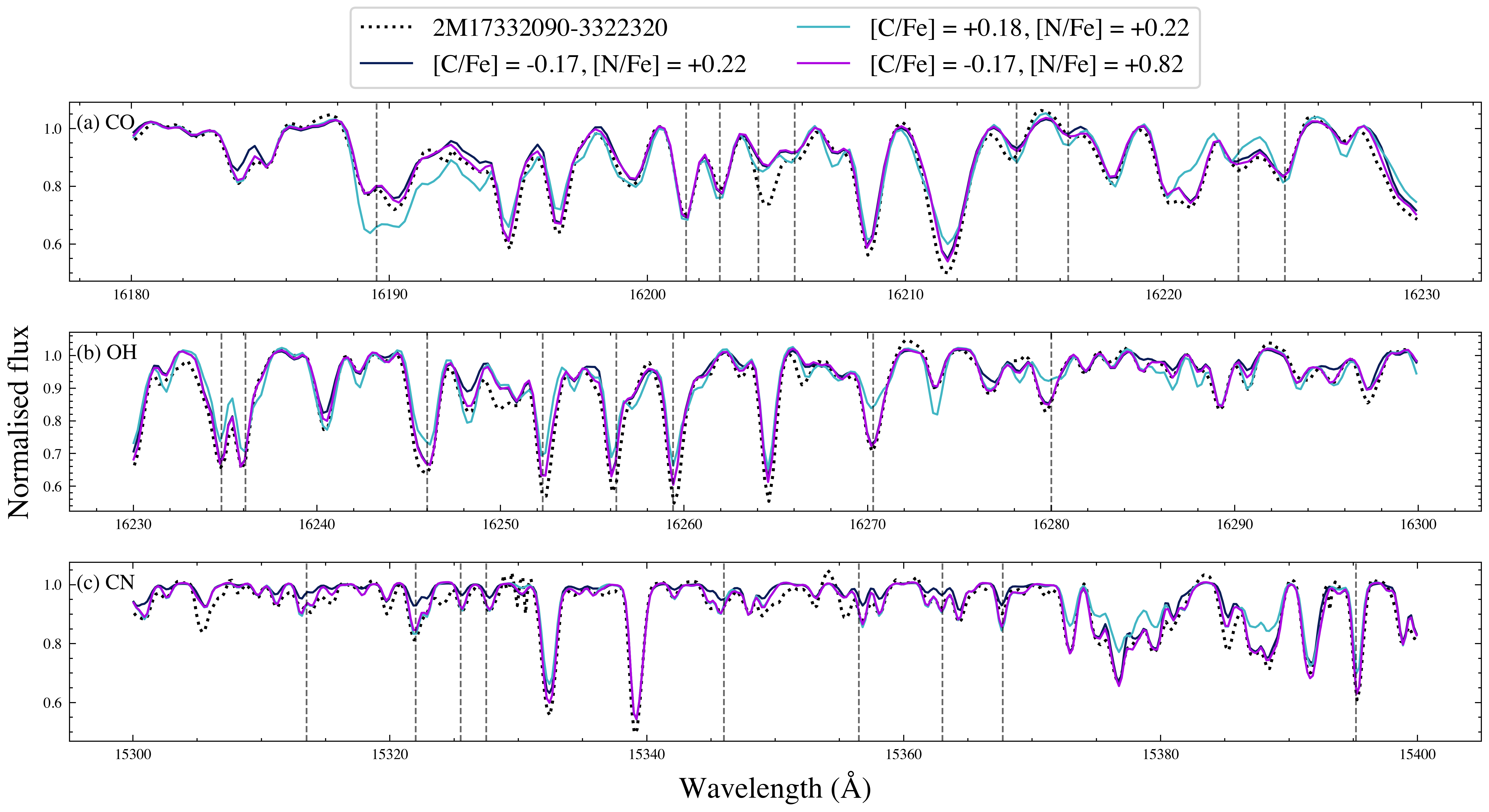}
    \caption{Combined APOGEE spectrum for star 2M17332090-3322320 (dotted line). The synthetic spectrum for the APOGEE abundances is shown in purple. Also plotted is the synthetic spectrum calculated assuming the same stellar parameters, for two different sets of abundance patterns. In one case C and N abundances adopted from a similar Liller 1 star whose C and N abundances are consistent with a 1G origin (navy line). In the other case, the spectrum is calculated adopting the same C abundance as that obtained by APOGEE for that star, but assuming a lower N abundance (cyan line).The panels (a), (b), (c) are the main CO, OH, CN molecular bands APOGEE uses to determine the respective abundances. The corresponding molecular lines are indicated with the grey dashed lines. All in all, the derived APOGEE abundances for this star are a good fit, confirming it is indeed rich in N, and thus characterised by a 2G abundance pattern.} 
    \label{fig:cnverify}
\end{figure*}

\begin{figure*}
    \includegraphics[width=1\textwidth]{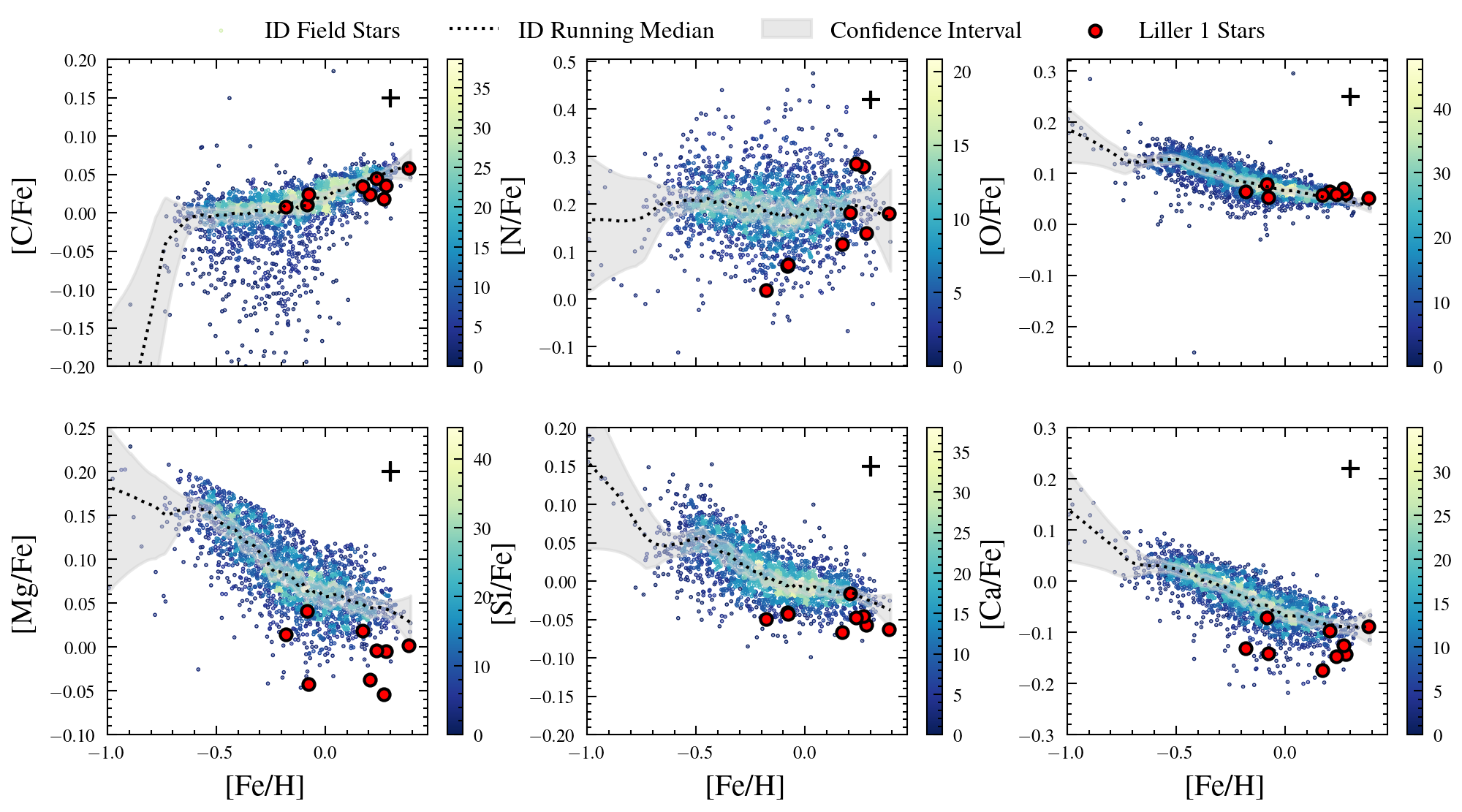}
    \caption{The [X/Fe]-[Fe/H] scatter plot with the number density of inner disk field stars described by the colour bar and Liller~1 stars shown in red. The respective running median is shown with the black dashed line.}
    \label{fig:lowessid}
\end{figure*}

\begin{figure*}
    \includegraphics[width=1\textwidth]{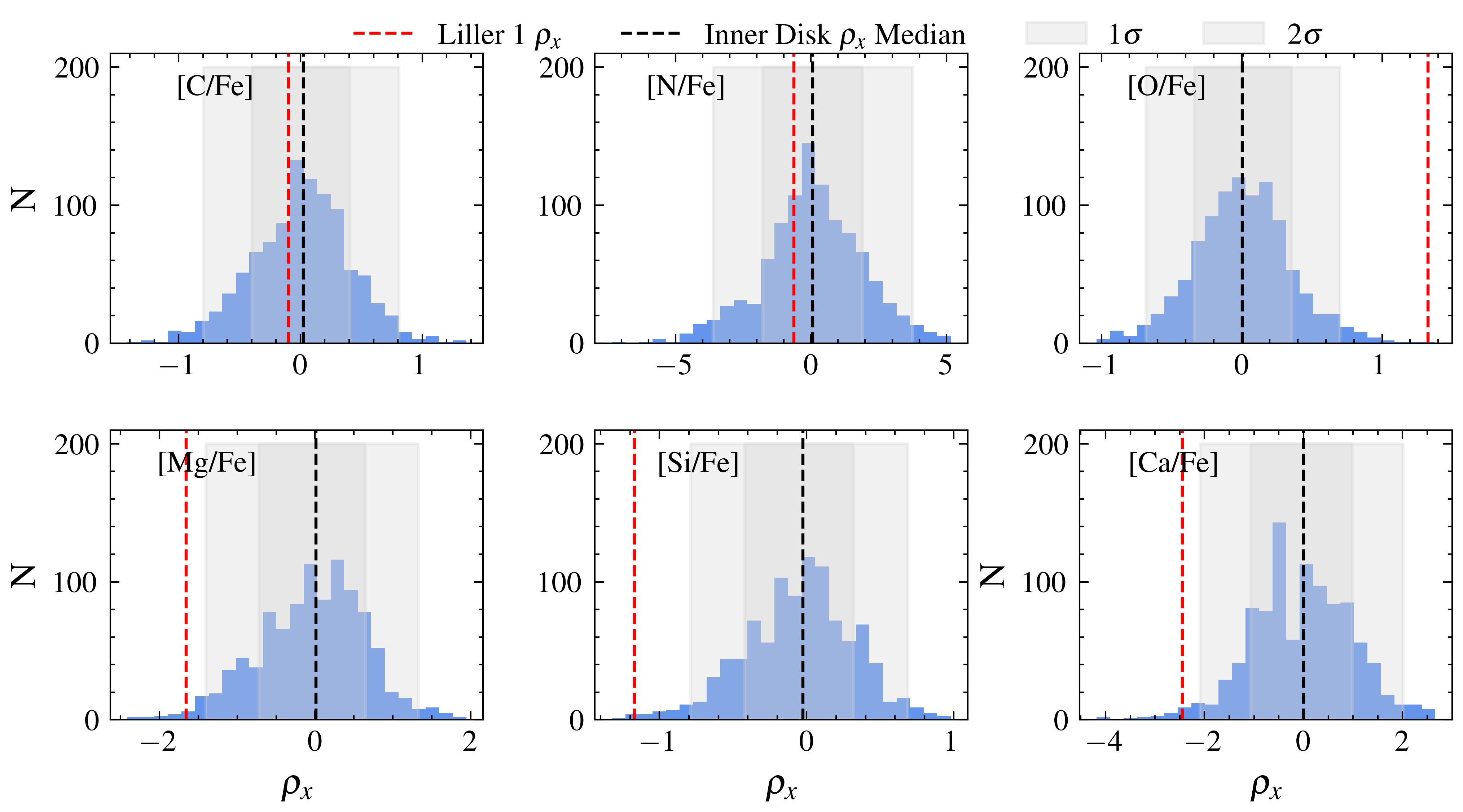}
    \caption{Histograms of the $\rho_X$ distributions of the inner disc random samples. The median of the randomly sampled distributed is indicated with the black dashed line. The light and dark grey shading indicate the $1\sigma$ and $2\sigma$ intervals respectively. The Liller~1 $\rho_X$ is indicated with the red dashed line.}
    \label{fig:idrho}
\end{figure*}

\begin{figure*}
    \includegraphics[width=1\textwidth]{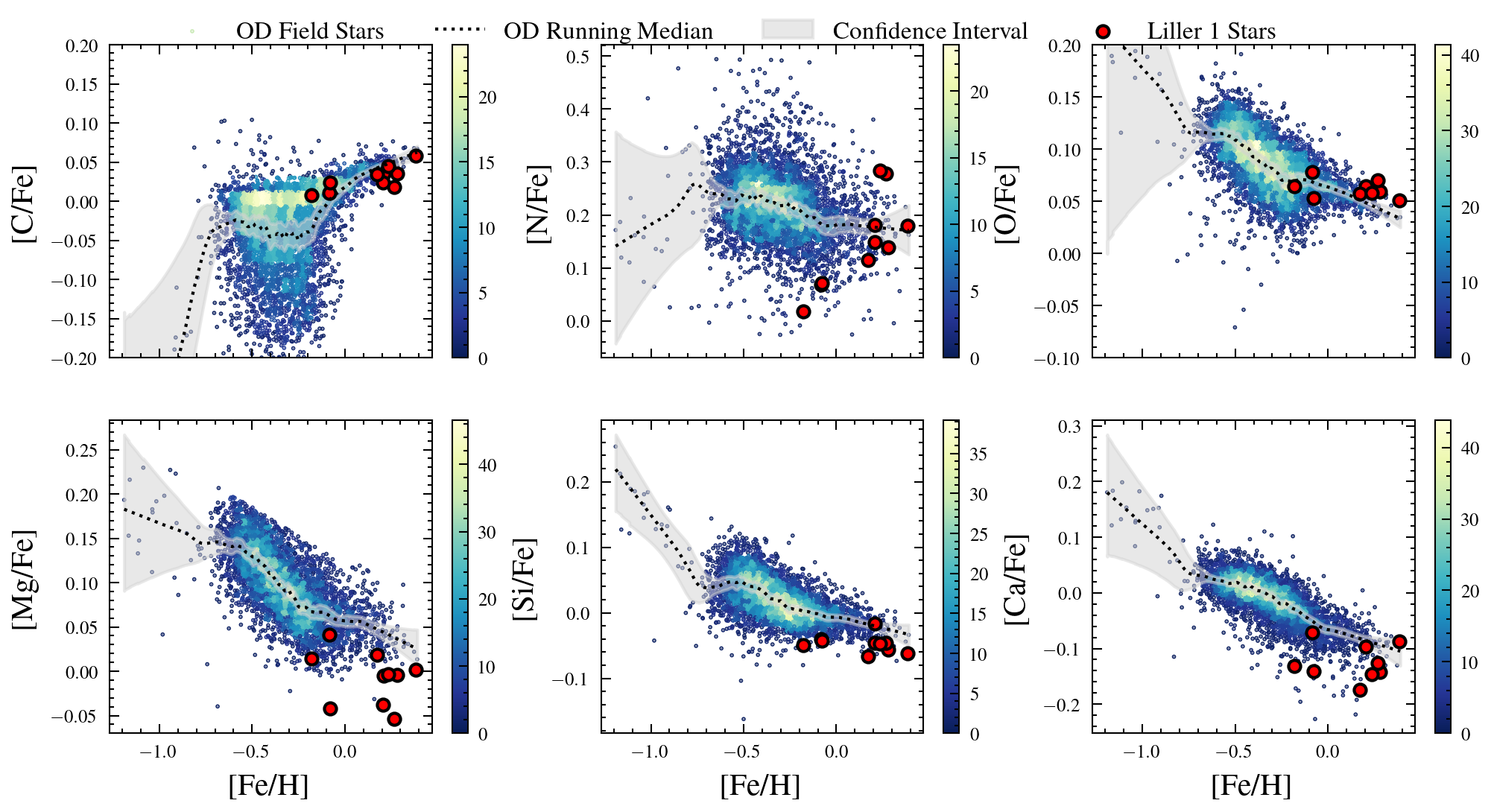}
    \caption{The [X/Fe]-[Fe/H] scatter plot with the number density of outer disk field stars described by the colour bar and Liller~1 stars shown in red. The respective running median is shown with the black dashed line.}
    \label{fig:lowessod}
\end{figure*}
\label{sec:Appendix}

\bsp 
\label{lastpage}
\end{document}